\documentclass[a4paper,11pt]{article}
\pdfoutput=1 

\usepackage{jinstpub} 
\usepackage{ulem}
\usepackage{xcolor}
\title{Test Beam Study of SiPM-on-Tile Configurations}

\author[b]{A. Belloni}
\author[b]{Y.M. Chen}
\author[a]{A. Dyshkant}
\author[b]{T. K. Edberg}
\author[b]{S. Eno}
\author[c]{J. Freeman}
\author[d,1]{M. Krohn\note{Corresponding author.}}
\author[b]{Y. Lai}
\author[c]{D. Lincoln}
\author[c]{S. Los}
\author[d]{J. Mans}
\author[d]{G. Reichenbach}
\author[c]{L. Uplegger}
\author[a]{S. A. Uzunyan}
\author[a]{V. Zutshi}

\affiliation[a]{Northern Illinois University}
\affiliation[b]{University of Maryland}
\affiliation[c]{Fermilab National Accelerator Laboratory}
\affiliation[d]{University of Minnesota}
\abstract{Light yield and spatial uniformity for a large variety of configurations of scintillator tiles were studied.  The light from each scintillator was collected by a Silicon Photomultiplier (SiPM) directly viewing the produced scintillation light (SiPM-on-tile technique). 
The varied parameters included tile transverse size, tile thickness, tile wrapping material, scintillator composition, and SiPM model. 
These studies were performed using 120 GeV protons at the  Fermilab Test Beam Facility. External tracking allowed the position of each proton penetrating a tile to be measured.  The results were compared to a GEANT4 simulation of each configuration of scinitillator, wrapping, and SiPM.}
\arxivnumber{2102.08499} 
\begin{document}
\rightline{Version 1.1,~~\today}
\maketitle
\flushbottom
\section{\label{sec:intro}Introduction}

The High Luminosity phase of the Large Hadron Collider (HL-LHC)~\cite{HLLHC} is scheduled to begin at CERN in 2027, with a designed instantaneous luminosity of $5\times 10^{34}\text{ cm}^{-2}\text{s}^{-1}$. In order to operate in this environment, 
a new high granularity calorimeter (HGCAL)~\cite{HGCAL} will be installed in the endcap regions of the CMS detector. 
In regions of the calorimeter where the fluence is highest, the design uses silicon sensors as the active material.  In lower fluence regions, the design uses plastic scintillator tiles, with the scintillation light readout by silicon photomultipliers (SiPMs).
%

The design of the scintillator section of HGCAL is driven by the necessity of calibrating the detector with minimum ionizing particles (MIPs) 
throughout the lifetime of the calorimeter.
To achieve the required performance, uniform light collection across the face of individual scintillator tiles is important. The SiPM-on-tile technology, where the SiPM is located in a dimple machined into the tile surface, 
has been studied by the Calorimeter for Linear Collider Experiment collaboration (CALICE)~\cite{CALICE1, CALICE2, TileOnSiPMtest}.  These studies reported that for $97\%$ of a $3\times3~\text{cm}^{2}$ tile area the light yield response is within $10\%$ of the average response. Similar studies are needed for the tile designs that will be used in the HGCAL. 
Some adjustment of the HGCAL SiPM-on-tile design may be required to cover the full range of tiles in the calorimeter. CMS HGCAL will use scintillator tiles of approximately square  shape varying in size from  roughly $2.3\times2.3$ to $5.5\times5.5~\text{cm}^{2}$.

We describe the apparatus and analysis used to study the responses of different scintillator 
tile geometries and materials using the Fermilab Test Beam Facility (FTBF)~\cite{FTBF} 
at the Fermi National Accelerator Laboratory. In this report, we present measurements for various geometries 
of Eljen Technology EJ-200~\cite{scinti_eff} and Kuraray SCSN-81~\cite{ref:kurary} tiles and their comparison with GEANT4~\cite{geant4-1, geant4-2, geant4-3} based simulations.
\section{Test beam setup}
The Fermilab Test Beam Facility~\cite{FTBF} provides a primary beam containing 120 GeV protons bunched at 53 MHz. 
The beam is delivered as a slow spill with a 4.2 second duration once per minute and intensity of approximately 
$5\times10^{4}$ protons per spill. The beam spot shape is roughly Gaussian, with standard deviation in x and y of about $1.5~\text{cm}$ (x and y are the horizontal and vertical directions transverse to the beam direction, respectively), 
to provide the sufficiently even population of protons across our scintillator tiles. 
We present the analysis of data taken during January-February 2020. 

\begin{figure}[htbp]
\centering 
\hspace*{1cm}
\includegraphics[width=1\textwidth]{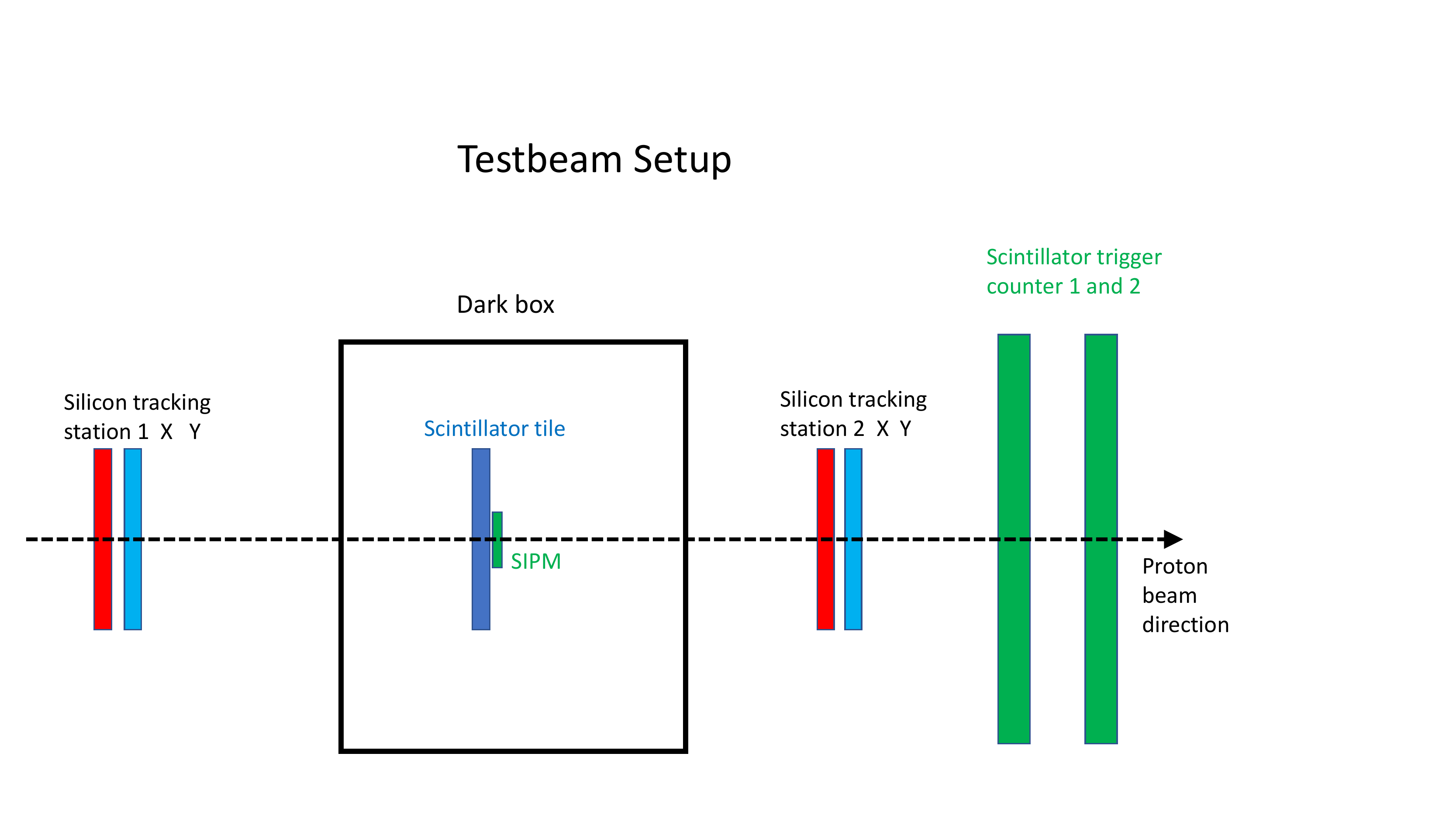}
\caption{\label{fig:testbeam setup} Test beam layout, not to scale. A dark box on a computer-controlled x-y stage contained the SiPM-on-tile sample.
Upstream and downstream doublets of silicon strip layers provided particle tracking. Two downstream scintillation counters provided the trigger.  }
\end{figure}

The test beam configuration is shown in Fig~\ref{fig:testbeam setup}. 
The scintillator sample and the readout SiPM were placed in a dark box mounted 
on a computer-controlled x-y stage that provided the ability to move the tile transversely to the beam. 
The readout SiPM was connected to a 2~GHz inverting amplifier~\cite{PORKA}. 
Bias voltage for the SiPM was provided by a Keithley 6410 source meter. 
The analog signal from the SiPM was sent to a DRS4 1 -- 5~GS/s waveform digitizer~\cite{PSI}.   
The firmware of the DRS4 was modified to provide an output "BUSY" signal~\cite{Paul Rubinov}.


Silicon strip tracking stations were placed on the upstream and downstream sides of the dark box containing the scintillator sample to measure the beam particle position. Two orthogonal layers of strips (tracker planes) in each tracking station recorded three-bit amplitude signals for each triggered event.  The strips were $60~\mu\text{m}\times10.0~\text{cm}$ and with a total of 640 per layer. The total overlap active area was $3.84\times3.84~\text{cm}^{2}$.

We used collections of strips with amplitudes above a threshold to find clusters of adjoining strips.  Amplitude-weighted coordinates of strips were assigned to these clusters to build particle tracks. Assuming parallel beam propagation, we treated (X, Y) pairs of clusters in the upstream and downstream stations with the minimum transverse distance between them as the most probable track trace. We used the (X, Y) position of the most probable track in the upstream tracker station as the beam particle's measured position (in events with no hits in the downstream tracker station, the first  (X, Y) pair of clusters in the upstream station was used).

The coincidence of two scintillator counters provided the trigger for the system. 
The coincidence was fed into a NIM+ module which is based on the CAPTAN+ board designed at Fermilab and a Kintex7 FPGA.  
This module was configured to accept triggers that occurred in the absence of BUSY signals from the DRS4 and the silicon readout, so that a single event was readout before a new trigger was sent.

\section{\label{sec:setup} DAQ setup}
The FTBF Data Acquisition System~\cite{OTSDAQ} is called OTSDAQ, "Off The Shelf Data Acquisition." 
It was designed at Fermilab and is based on the XDAQ libraries developed at CERN for the CMS experiment. 
OTSDAQ was used to configure the DRS4 board and the silicon strip tracker stations, 
to provide triggers for both systems, and to perform online data quality monitoring and run control. 
The digitized waveforms from the DRS4 and the tracker information data streams were collected by the independent DAQ computers.  
Offline, these streams were combined into events using matching triggers. 
The operating acquisition rate was about 1K events per beam spill.
\section{\label{sec:concept}SiPM-on-Tile Concept}
This study investigates prototypes based on the SiPM-on-Tile concept~\cite{CALICE1}. 
Scintillators of various sizes were machined with a "dimple" as shown in Fig~\ref{fig:tile design}. 
The dimple provided a physical space for the SiPM and improved uniformity of response. 
In general, the SiPM will detect more light from a particle passing close to it. To reduce this geometry-induced hot spot,
a dimple was machined to reduce the amount of scintillator and hence the amount of light generated near the SiPM.
The SiPMs were located at the center of the dimple, with the active face located $0.55~\text{mm}$ 
into the dimple relative to the tile face. A reflective wrapper surrounded the tile 
to reflect escaping photons back into the tile, increasing the light yield. 
The wrapper covered 100\% of the tile area except for a small hole at the dimple to accommodate the SiPM. 
%
%
%
%
\subsection{Description of tile geometries}
\begin{figure}[htbp]
\centering 
\includegraphics[width=.45\textwidth]{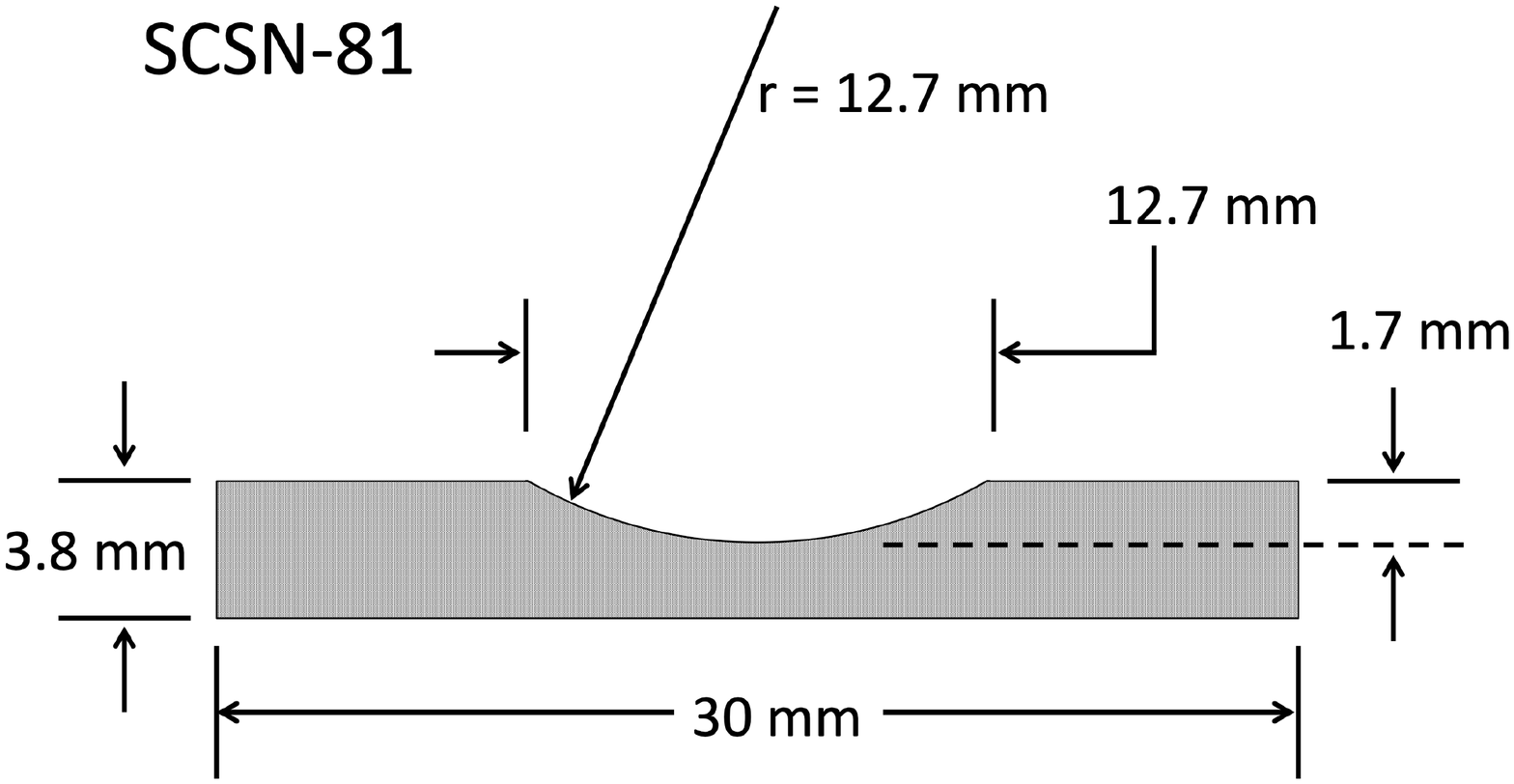}
\qquad
\includegraphics[width=.45\textwidth]{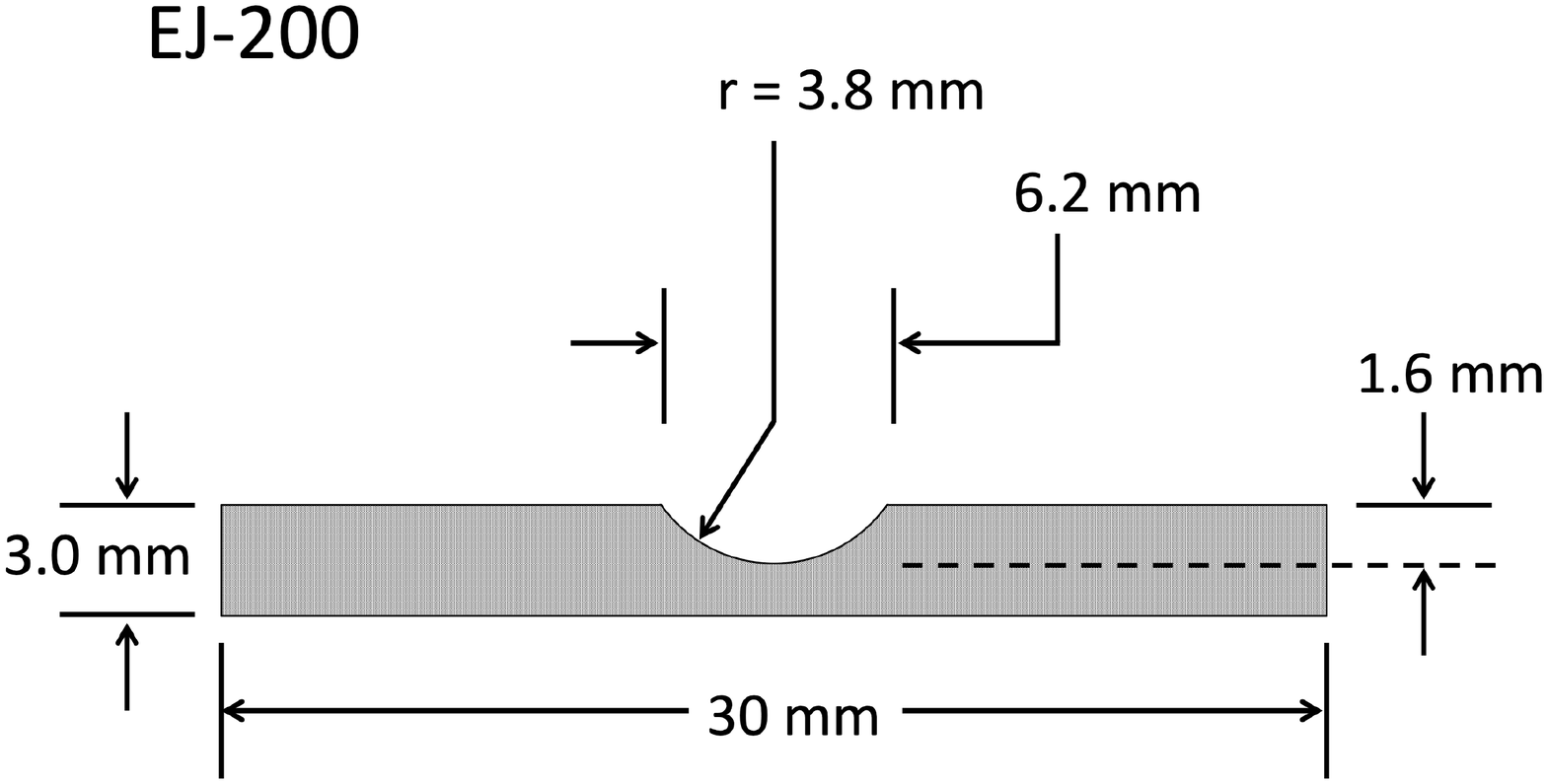}
\caption{\label{fig:tile design} Left: Cross section of "SiPM-on-Tile" SCSN-81 Design. Right: Cross section of EJ-200 tile. The dimples are centered on the tile face. }
\end{figure}

Tiles were prepared from two scintillator materials, $3.8~\text{mm}$ thick SCSN-81 (Kuraray), 
and $3~\text{mm}$ thick EJ-200 (Eljen Technology).  Dimples were machined into a flat face of the tiles. 
Two different diameters of dimple were used, as shown in Fig~\ref{fig:tile design}: The SCSN-81 tiles had a spherical cap dimple  with a diameter 
on the face of $12.7~\text{mm}$ and depth of $1.7~\text{mm}$, while EJ-200 tiles had a smaller spherical cap dimple of $6.2~\text{mm}$ diameter on the face and depth of $1.6~\text{mm}$.
Tiles of various transverse sizes were prepared, ranging from  $2.3\times2.3$ to $5.5\times5.5~\text{cm}^{2}$.
\subsection{SiPMs used for the study}
Two SiPMs from Hamamatsu Photonics were tested: S13360-1350PE~\cite{S13360} and  S14160-1315PS~\cite{S14160}. 
These both have an active area $1.3\times1.3~\text{mm}^{2}$. The S13360-1350 has 50 micron pixels, 
while the S14160-13115 has 15 micron pixels. The gain of
the  S13360 is consequently about $10$ times larger. To ensure the pulse height of the two SiPMs were approximately the same, 
a second $10\times$ 2~GHz amplifier was employed when studying the S14160. 

\vspace{3mm} 
\section{Simulation}
\label{sec:simulation}
The expected optical responses of the various tiles were simulated using the GEANT4~\cite{geant4-1, geant4-2, geant4-3} toolkit.
The results are compared with data in Section~\ref{sec:response}.

\begin{figure}[htbp]
\centering 
\includegraphics[width=.4\textwidth]{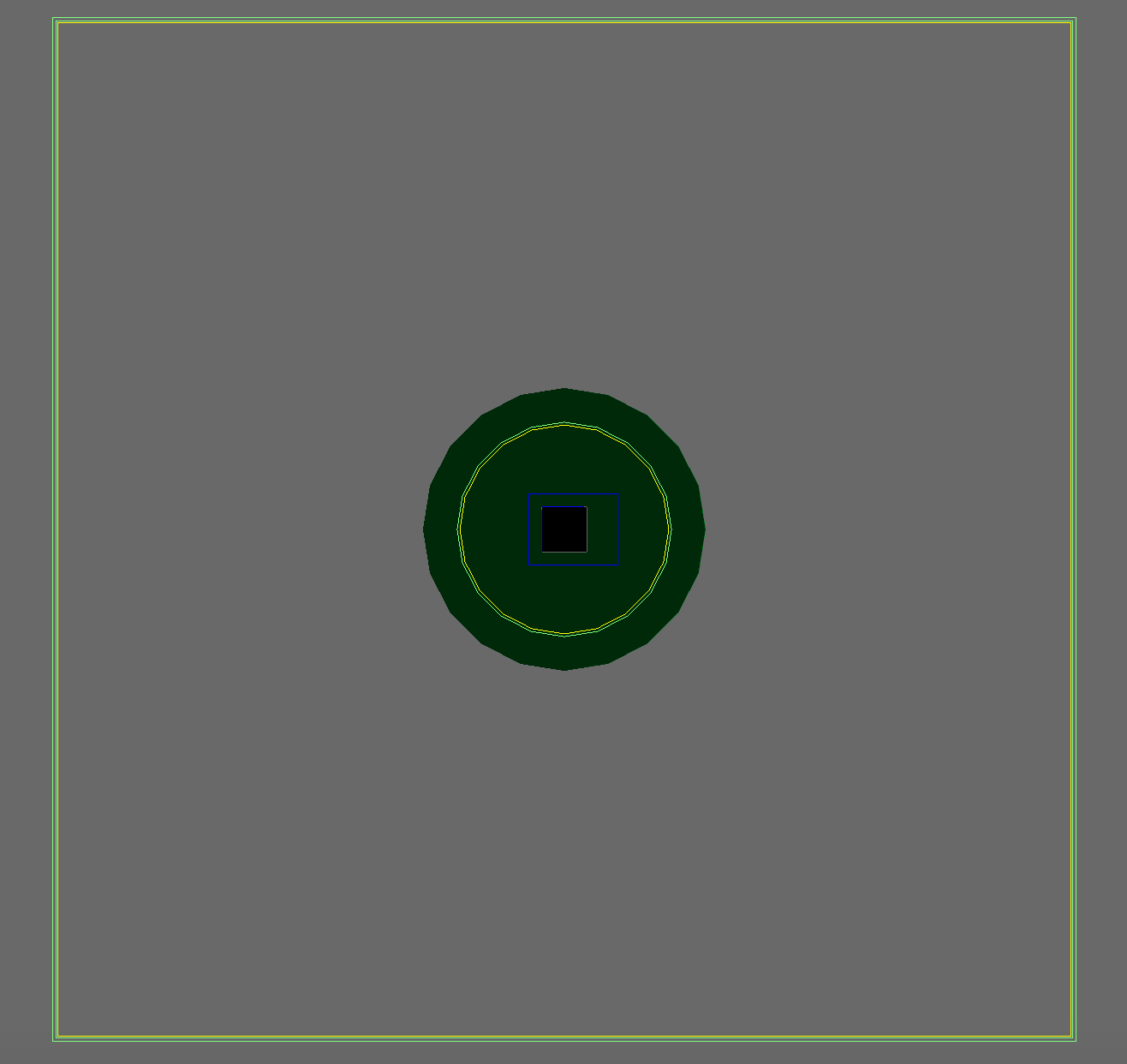}
\qquad
\includegraphics[height=.4\textwidth]{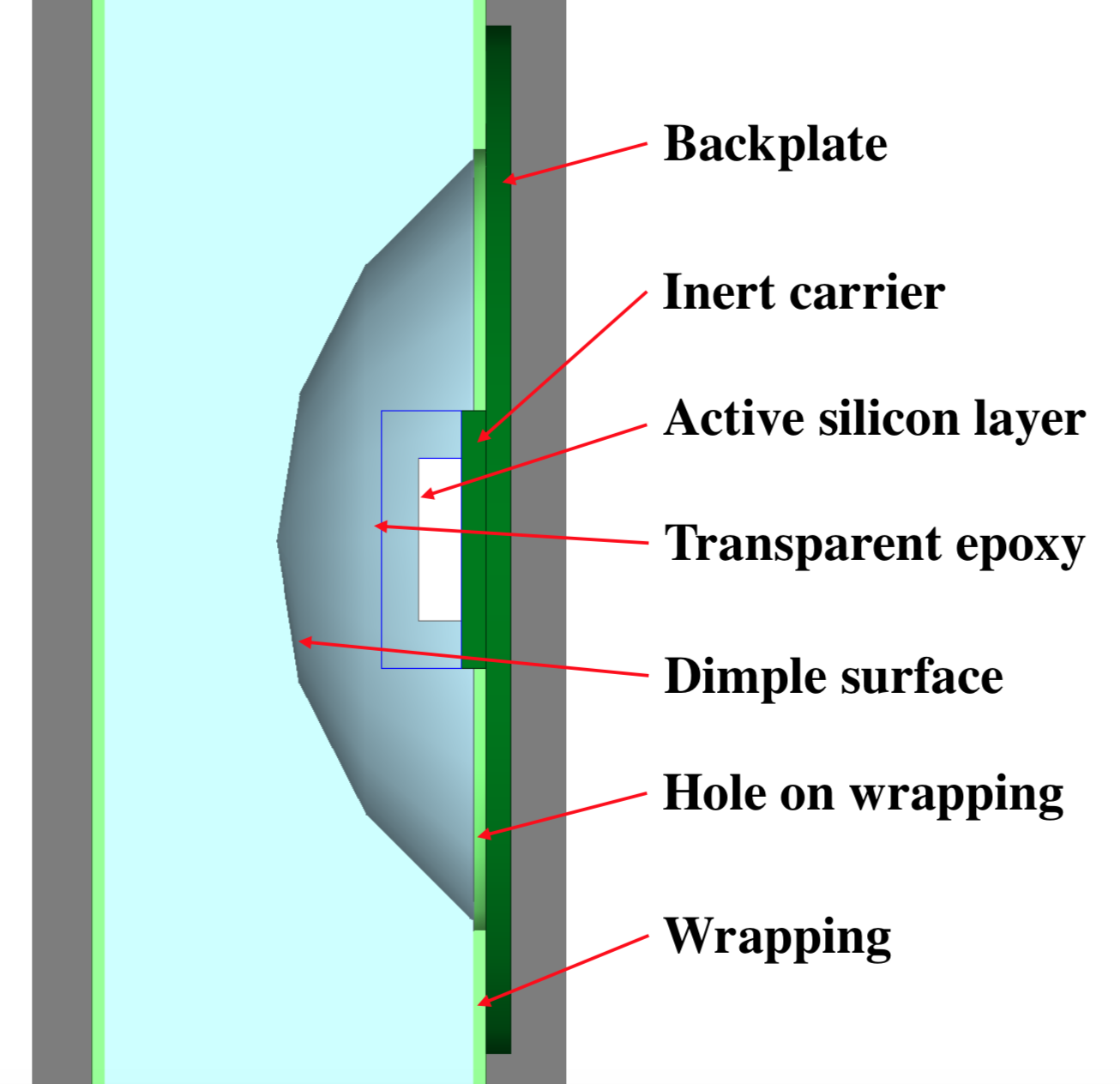}
\caption{\label{fig:g4simu} GEANT4-rendering of simulated tile geometry as seen by the beam (left) and side view (right).  
The tile is a rectangle composed of plastic scintillator with a spherical dimple.  
The square at the center of the dimple is the rendering of the simulated SiPM.
The sizes of this geometry follows Fig.~\ref{fig:tile design}~(right).
}
\end{figure}

Fig.~\ref{fig:g4simu} shows the GEANT4 rendering of the simulation geometry. 
It is separated into three parts:
the tile with its dimple, the reflective wrapping, and the SiPM attached to a backplate. 
The tile is completely wrapped in a reflective coating except for a hole centered on the dimple.
For most simulations, the SiPM back was flush with
the tile surface.
A circular backplate
outside of the wrapping was attached to the back of the surface-mount SiPM and was simulated as either printed circuit board white silk screen (WSS) or black tape with reflectivities at a wavelength of $425~\text{nm}$ of 0.68 and 0.05, respectively. 

 The simulated scintillator material was set to be the base material for SCSN-81 (polystyrene) and EJ-200 (polyvinyltolulene).  
Its optical properties were taken from the datasheets of the commercial products SCSN-81~\cite{ref:kurary} and  EJ-200~\cite{scinti_eff} except for the absorption length of EJ-200. 
The effective light attenuation length at $425\text{\,nm}$ was set to $3.8\text{\,m}$ for EJ-200.
The wavelength dependence of the attenuation length for EJ-200 was determined using a Varian Cary~300 spectrophotometer.  Varying the spectral dependence of the attenuation length within measured uncertainties made negligible differences in our simulations.
The attenuation length for SCSN-81 was set to be constant, $1.4\text{\,m}$ ~\cite{ref:kurary}.
%

The SiPM was modeled as a rectangular solid divided into three layers: transparent epoxy, an active silicon layer, and an inert carrier. 
Its detection efficiency and specification 
were taken from the Hamamatsu S13360-1350PE ~\cite{S13360} and S14160-1315PS~\cite{S14160} product data sheets. Only photons that hit the front face of the SiPM were recorded. The reflectivities of the inert carrier were set to be 0.5. 

The tile was modeled to have a perfectly polished surface, where reflection and refraction angles were deterministic.  Adding plausible surface imperfections had little effect on our results.
Two wrapping materials were simulated, Tyvek\textsuperscript{\textregistered} and 3M Vikuiti\texttrademark ~Enhanced Specular Reflector film (ESR)~\cite{wrapping} 
with effective reflectivities set to 0.79 and 0.985 at 425 nm, respectively. 
A thin air gap between the tile and the wrapping was included. 

A $120~\text{GeV}$ proton beam was generated, oriented 
perpendicular to the scintillator face.
The scintillating efficiency was $10 ~\text{photons}/\text{\,keV}$ for EJ-200~\cite{scinti_eff} and $8.7 ~\text{photons}/\text{\,keV}$ 
for SCSN-81~\cite{ref:kurary}.
A photon entering into the SiPM sensitive area was counted if it passed an acceptance-rejection cut on the SiPM photon detection efficiency curve. 
The average number of detected photons for each geometry was used as a metric for the simulated light yield.

For each geometry, the simulated proton's impact positions were uniformly distributed across the tile face.
We studied the effect on the light yield of a uniform beam and a Gaussian beam, with a width similar to the test beam, and found the difference to be less than  2\%.

When simulations were compared with the data in Section~\ref{sec:response}, an additional normalization constant was used to scale down the predicted light yield from simulation. It was $1.15$ and $1.01$ for EJ-200 and SCSN-81, respectively. These quantities took into account the difference between the most probable value (MPV) of light yield from data and the mean light yield from simulation and the estimations of the effective reflectivities that were used for the wrappings.

\section{SiPM calibration}
\begin{figure}[ht]
\centering
{
\includegraphics[scale=0.37]{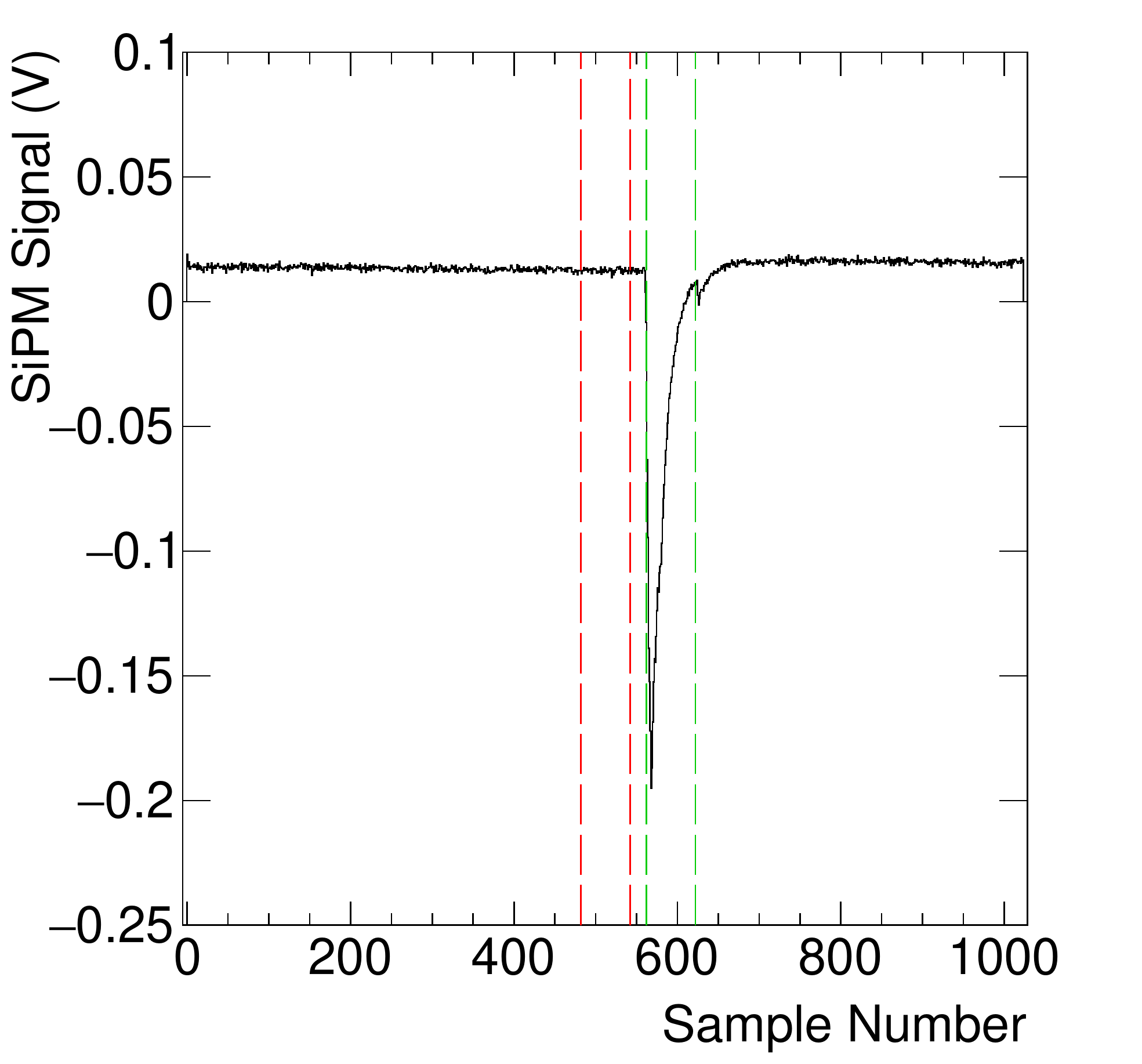}
\includegraphics[scale=0.37]{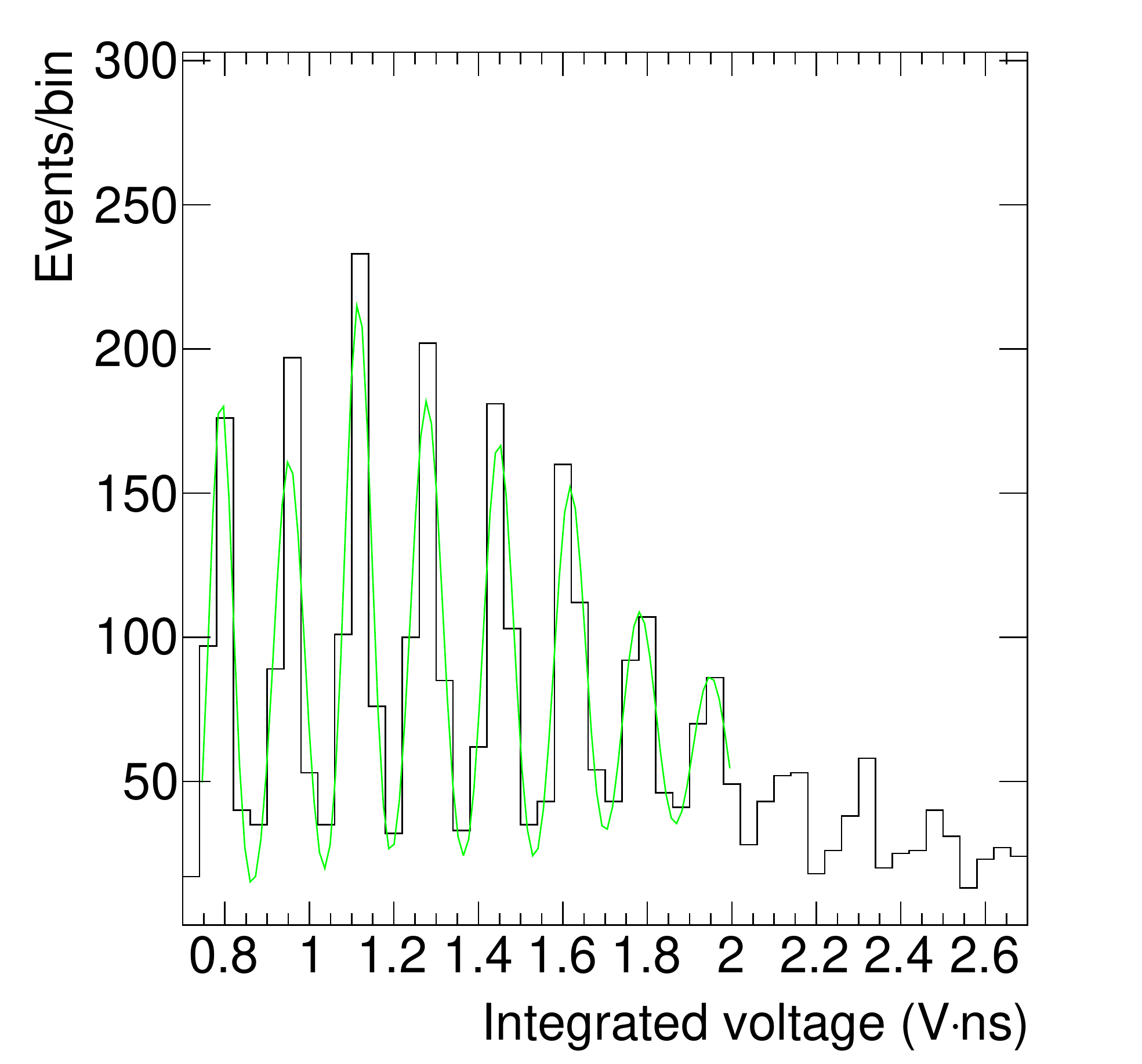}
\leftline{ \hspace{3.7cm}{\bf (a)} \hfill\hspace{3.7cm} {\bf (b)} \hfill}
\caption{\label{fig:calib1}(a) A typical SiPM signal waveform  ($1~\text{ns}$ per sample) and (b) a photoelectron spectrum  with multiple Gaussians fit over the peaks. 
The signal and pedestal regions of the waveform are integrated in regions [562, 622] (signal, green) and  [482, 542] (pedestal, red). }
}
\end{figure}

The DRS4 digitizes the SiPM analog voltage output using a 14 bit ADC with a configurable sampling frequency.  For this experiment, we operated the DRS4 at 1~GS/s.
For each trigger the SiPM waveform (as shown for example in Figure~\ref{fig:calib1}a) was recorded with 1024 samples.  
The DRS4 trigger delay was set to provide the pre-signal region of the waveform for the pedestal evaluation.   
We used integrals of  waveform pulses to measure the light output of the scintillator tiles. 
The waveform pulse was defined as a contiguous set of samples near a local maximum with an ADC amplitude above a threshold of $10~\text{mV}$.
Pulse voltage integrals  $V_I$ were calculated as the sums of waveform amplitudes in a region of 60~samples, 
starting with the sample with a signal above 0.25 of those at a pulse maximum.  To obtain the conversion factor between the 
integrated voltage and the number of the photoelectrons (PEs) produced in the SiPM, the following procedure was applied:

\begin{itemize}
\item  we collected 10-20K  waveforms  for  a scintillator  tile with a known low light output per incident proton in order to see PE peaks;
\item  for each waveform, we calculated the signal integrated voltage $V_I$ under the pulse with maximum amplitude;     
\item  the pedestal integrated voltage was calculated using a sample region of the signal pulse width in the pre-signal part of the waveform; 
\item  after subtraction of the pedestal, signals were histogrammed to obtain the PE spectrum;
\item  a fit (the sum of Gaussian curves) of the first six peaks of the spectra was used to calculate the conversion factor as the mean distance between peaks in the fit region. 
\end{itemize}

Figure~\ref{fig:calib1}a shows a DRS4 waveform collected from a S14160-1315 SiPM, when a 120~GeV proton passes through a Tyvek\textsuperscript{\textregistered}-wrapped SCSN-81 scintillator tile.  
The difference of integrals in regions [562, 622] (signal) and  [482, 542] (pedestal) is taken for the signal histogram in this event.  

Figure~\ref{fig:calib1}b shows a fitted spectrum with individual PE peaks corresponding to a calibration factor of  $0.165\pm 0.001~PE/V.$ 
Given the number of pixels in the SiPMs and the photon detection efficiency of the SiPMs at the voltages we operated them, the SiPMs response was linear over the PE ranges studied.
Numerous calibration measurements were made throughout the data taking to account for any drift in the calibration factor due to temperature or other external factors.

\section{\label{sec:response}Scintillator response}
\subsection{Event selection}
We selected events which 1)~have a matching trigger number for both DRS4 and
tracker data, and 2)~passed quality selection criteria.  We required:

\begin{itemize}
\item  a clean waveform -  the falling edge of the signal pulse was required to reach the level of 0.25 of the pulse's maximum 
and the pre-signal region should be wider than the number of samples used for signal integration;  
\item  noise suppression   -   SiPM signal pulses were required to have the peak amplitude above ten mV;
\item  clean tracker data  -  the upstream tracker station was required to have silicon strips fired in both x and y planes but have not more than one two-strip cluster in each. And the hit in the downstream tracker was required to be within $2~\text{mm}$ of the upstream hit in the direction transverse to the beam.
\end{itemize}

These cuts were sufficient to ensure that a single particle passed through the tile and that the particle can be treated as a MIP. 
Light yield distributions for the $3\times3~\text{cm}^{2}$ SCSN-81 tile wrapped in ESR 
before and after quality selections are shown in Figure~\ref{fig:evtsel}a. 
\subsection{The light yield determination } 
\begin{figure}[ht]
\centering
{
\includegraphics[scale=0.37]{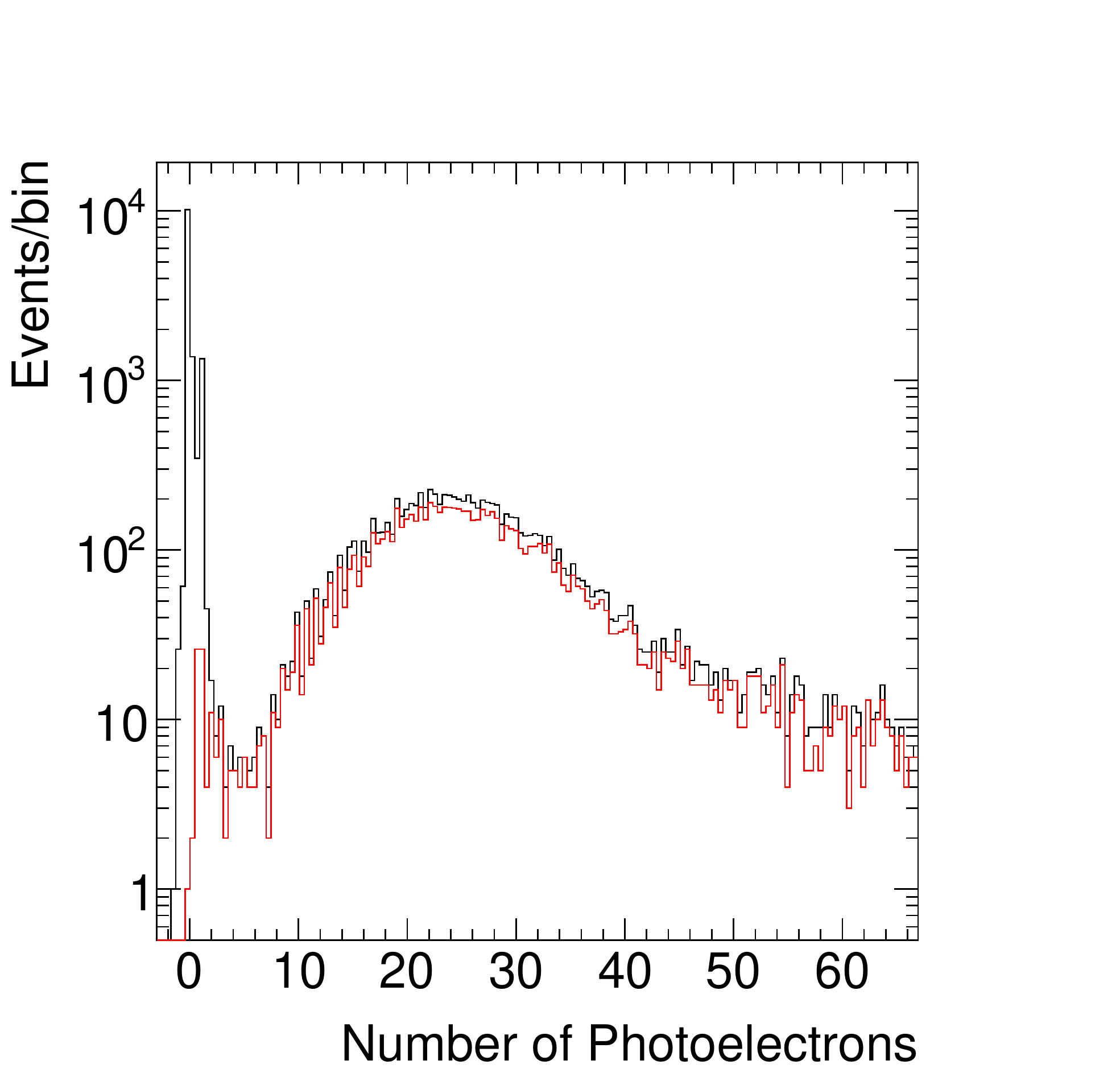}
\includegraphics[scale=0.37]{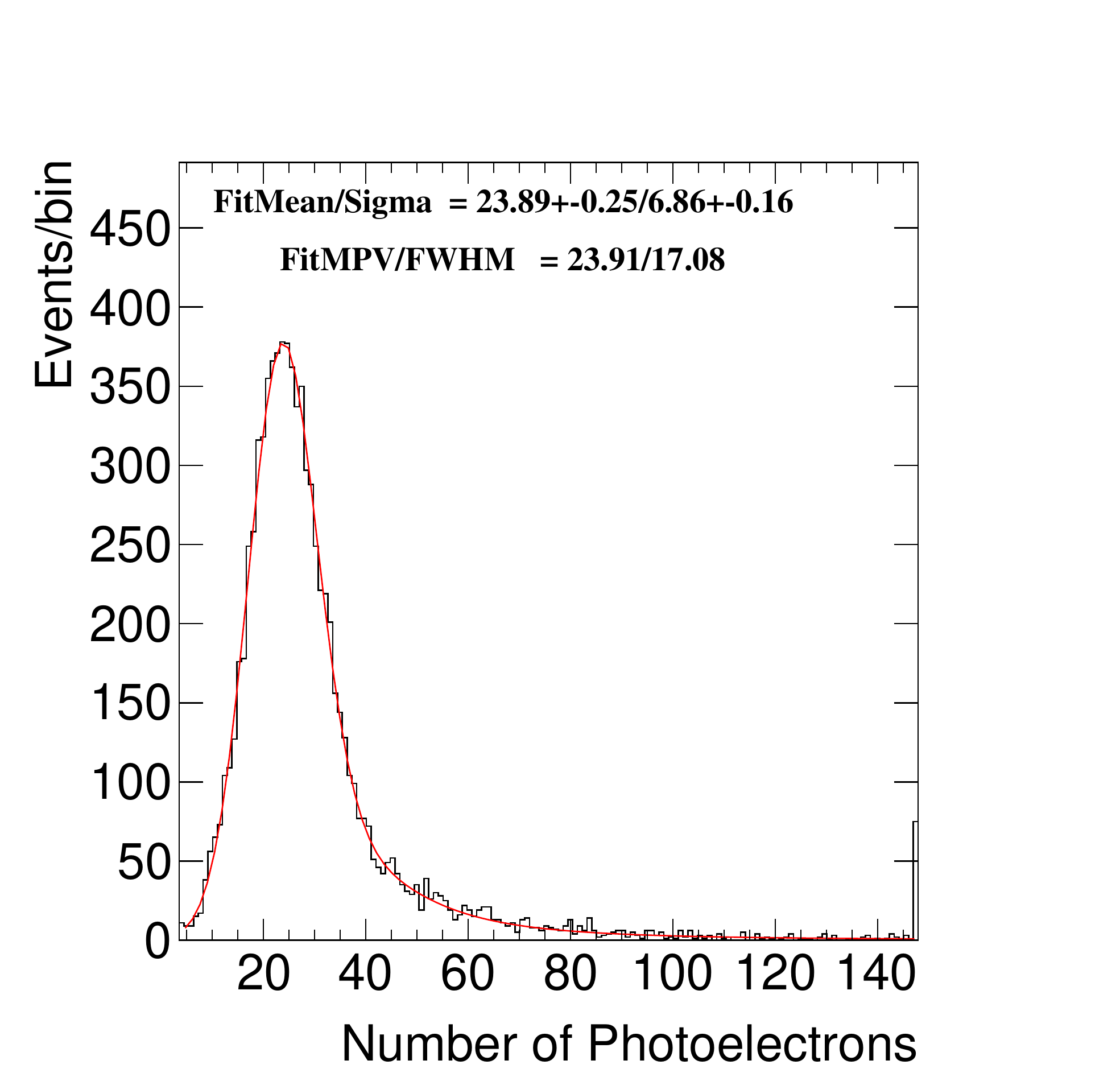}
\leftline{ \hspace{3.7cm}{\bf (a)} \hfill\hspace{3.7cm} {\bf (b)} \hfill}
\caption{\label{fig:evtsel} 
 (a) light yield distributions for the $3\times3~\text{cm}^{2}$ SCSN-81 tile wrapped in ESR before (black) and after quality selections (red), and (b) the fit of the PE yield distribution after quality selections.}
}
\end{figure}

To determine the light yield of a scintillator tile, the light yield distribution is fit with the sum of a Gaussian and a Landau function, where the lower light yield region is modelled to follow Gaussian statistics, but the high light yield tail is modelled with a Landau. 
We report the light yield as the most probable value (MPV) of the fit function.
We observed  that the MPV statistical uncertainty in a typical single measurement (5-10K clean events)
was smaller than the $\pm3\%$ optical coupling systematic uncertainty described below in Sec~\ref{sec:OpticalCoupling}. We also note that SiPM optical cross talk corrections to the light yield were not made. The cross talk varied for each type of SiPM and was subsumed in the systematic error quoted.  
Figure~\ref{fig:evtsel}b shows the fit of the light yield distribution for clean events in the $3\times3~\text{cm}^{2}$ SCSN-81 tile test. 
\subsection{Tile uniformity}

\begin{figure}[ht]
\centering
{
\includegraphics[width=.7\textwidth]{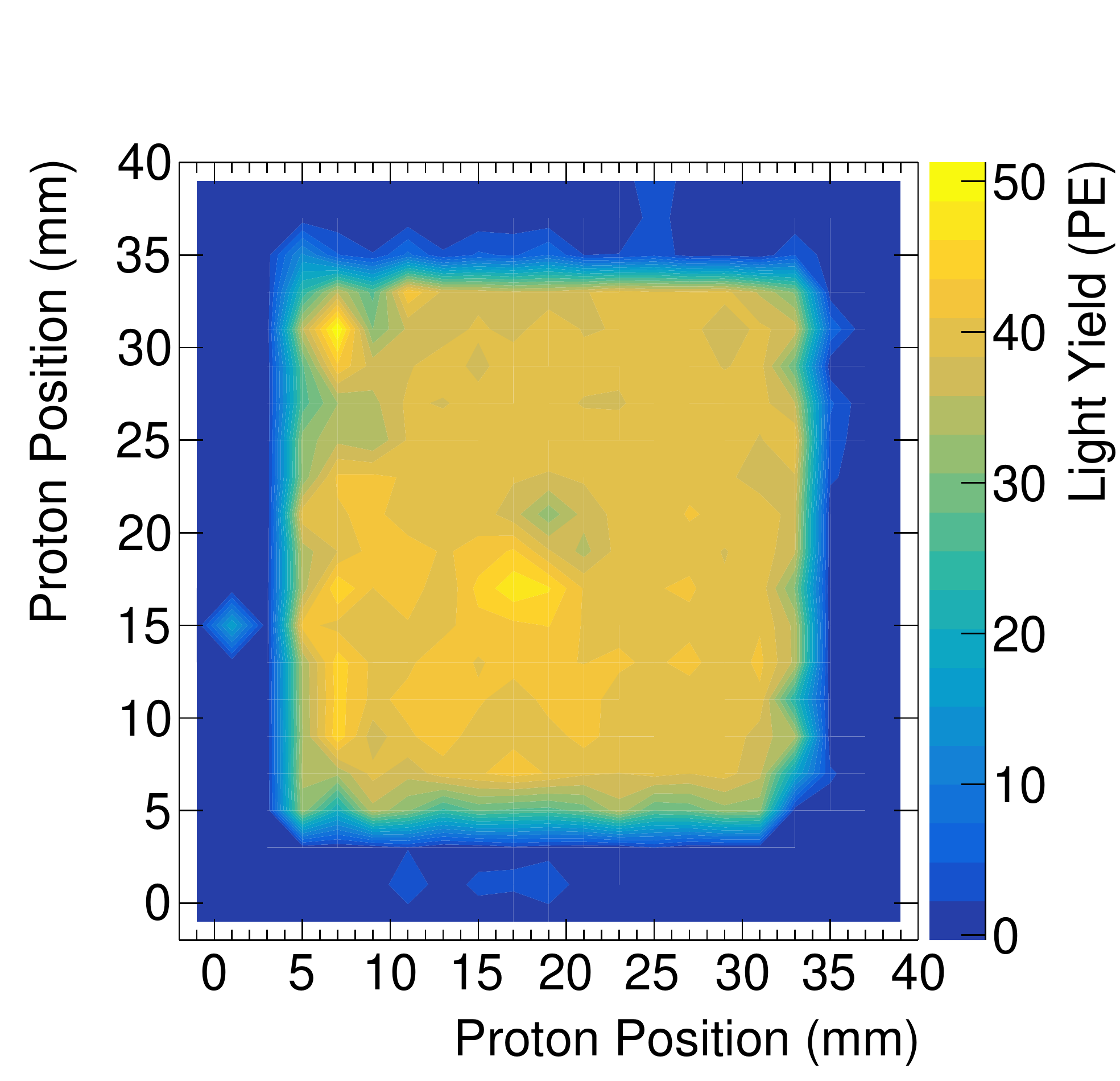}
\caption{\label{fig:EJ200_uniformity} 
The average light yield distribution of the EJ-200 reference tile as a function of proton position in the upstream tracker.}
}
\end{figure}

\begin{figure}[ht]
\centering
{
\includegraphics[scale=.37]{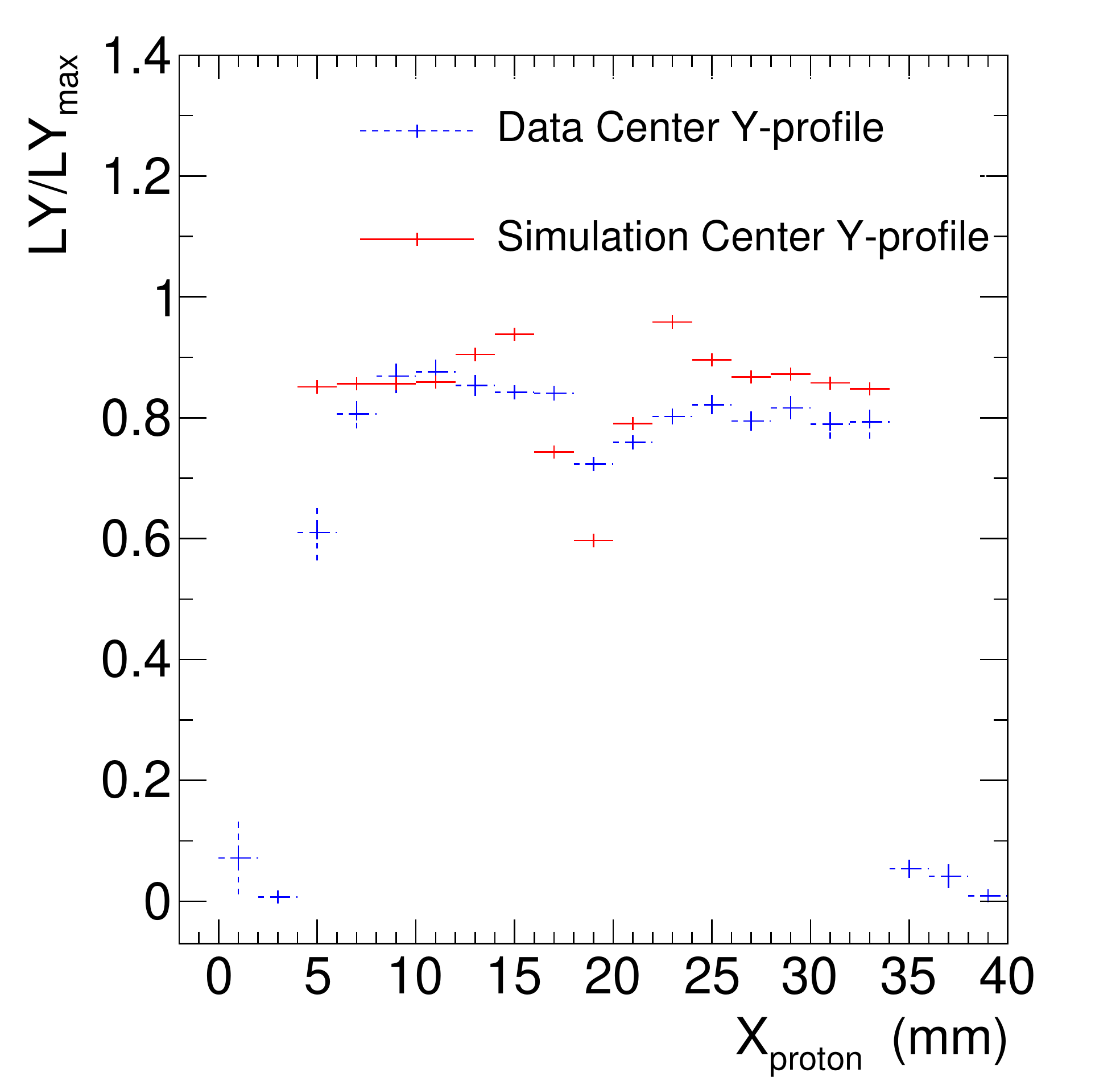}
\includegraphics[scale=.37]{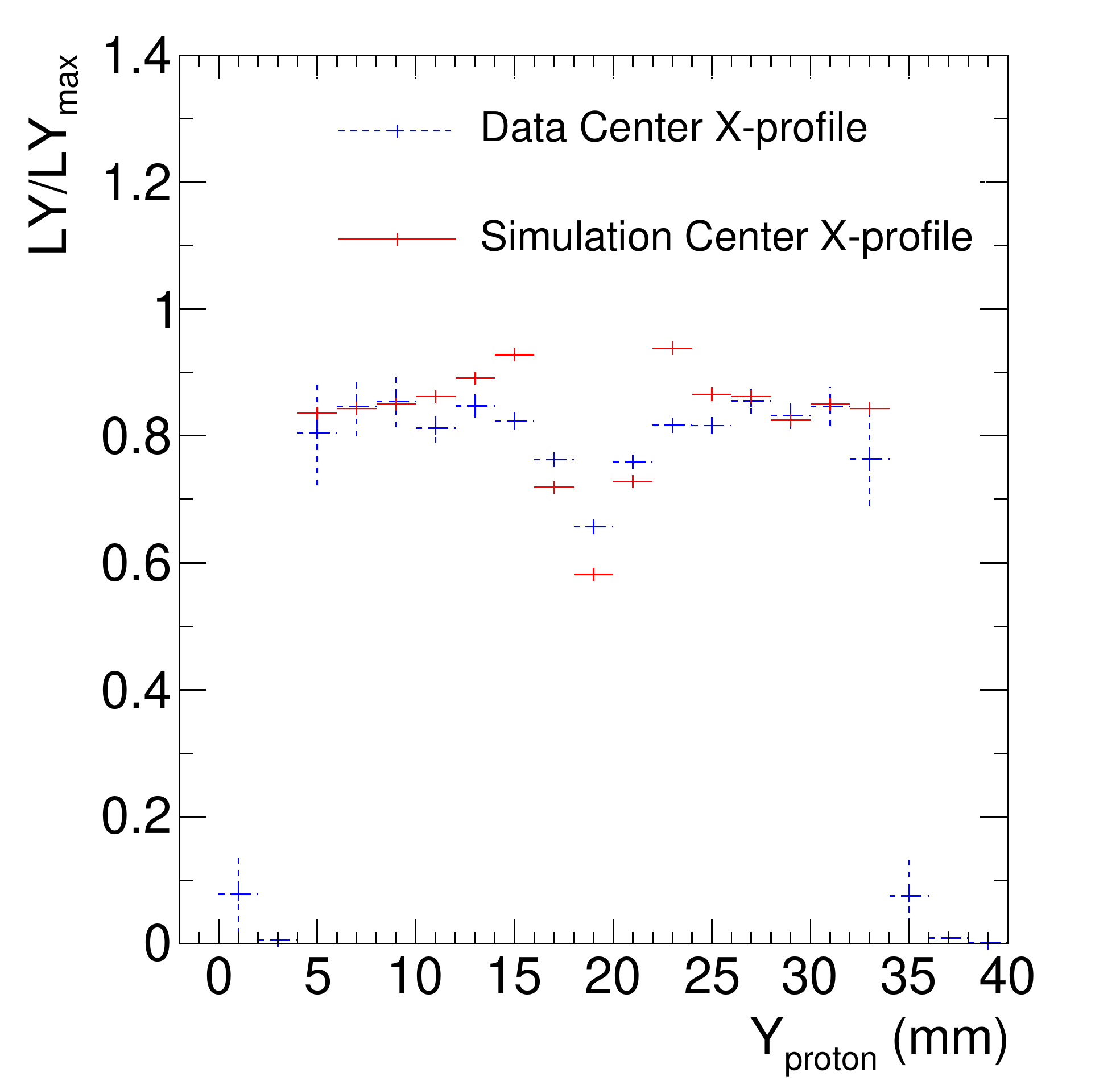}
\leftline{ \hspace{3.7cm}{\bf (a)} \hfill\hspace{3.7cm} {\bf (b)} \hfill}

\includegraphics[scale=.37]{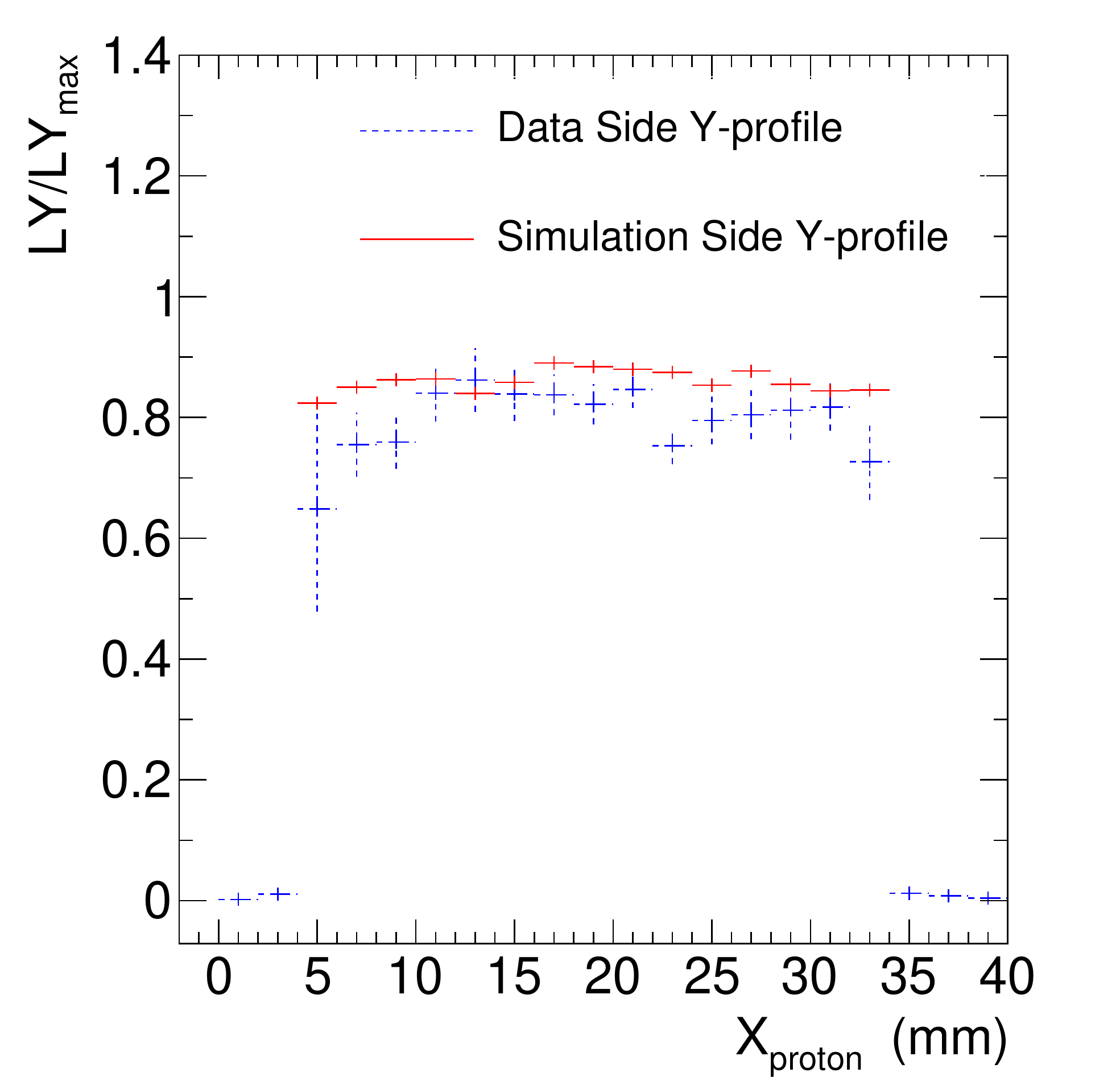}
\includegraphics[scale=.37]{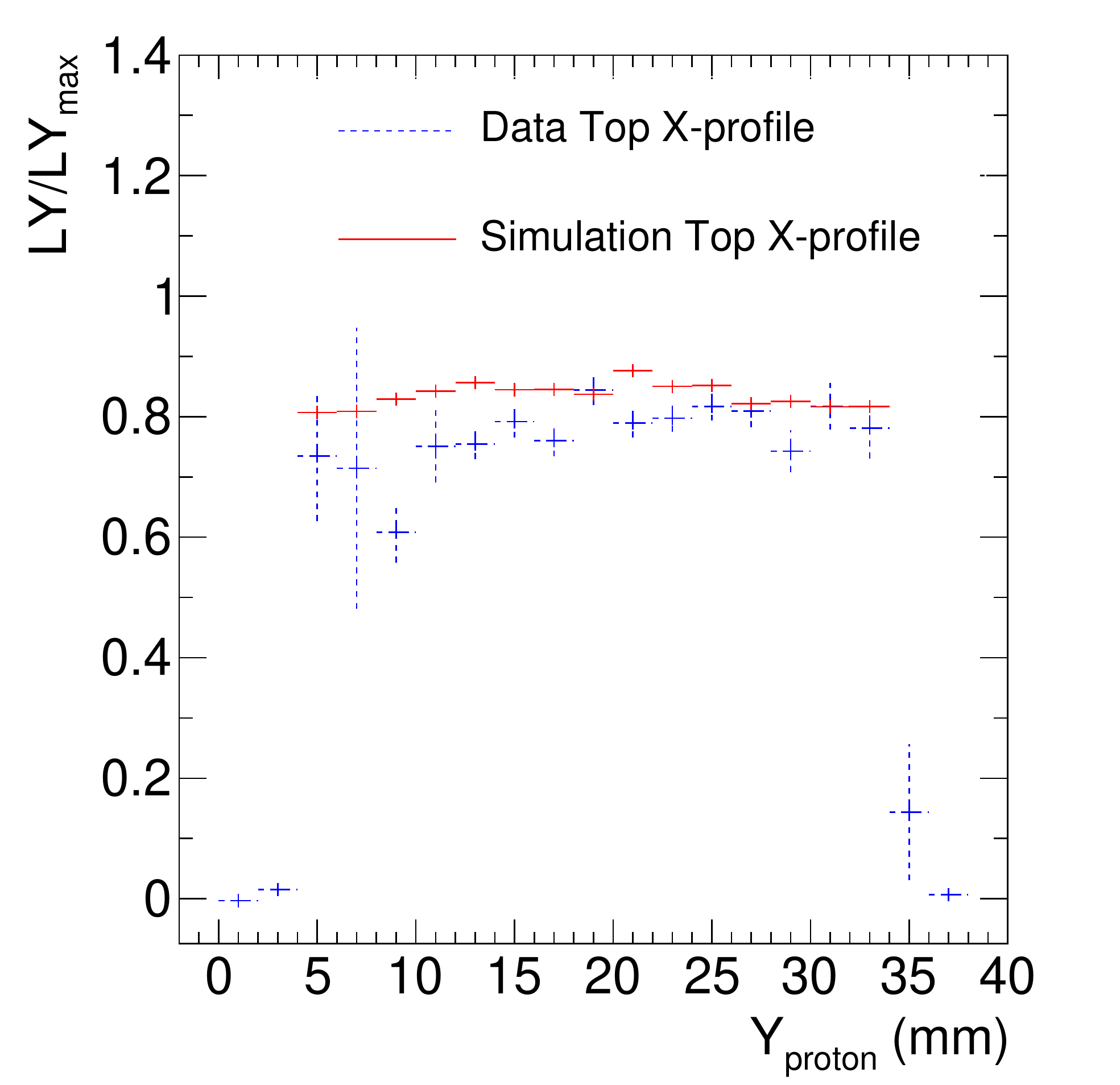}
\leftline{ \hspace{3.7cm}{\bf (c)} \hfill\hspace{3.7cm} {\bf (d)} \hfill}
\caption{\label{fig:uniformity_profilesEJ200} 
 Normalized light yield distribution profiles in data and simulation for the EJ-200 reference tile along 
 a) $2~\text{mm}$ y-bands at the tile center, b) $2~\text{mm}$ x-bands at the tile center,
 c) $2~\text{mm}$ y-bands offset by $8~\text{mm}$ in the x direction from the tile center, and 
 d) $2~\text{mm}$ x-bands offset by $8~\text{mm}$ in the y direction from the tile center.
 In the top distributions, the sharp peaks in simulation correspond to the edge of the dimple and this feature
 does not appear in the data due to the machining of the dimple.}
}
\end{figure}

An attractive feature of the experimental setup is the capability for studies of the spatial uniformity of the collected signals across the face of a tile.  
Fig.~\ref{fig:EJ200_uniformity} shows the average light yield distribution for the $3\times3\times0.3~\text{cm}^{3}$ EJ-200 tile wrapped in ESR (reference tile) as a function of the proton impact position. The observed pattern in the light yield, including a bright spot slightly below the dimple region, is consistent with a slightly misaligned SiPM as reported in~\cite{ref:asymmetry}.
Fig.~\ref{fig:uniformity_profilesEJ200} shows profiles of light yield distributions in data and simulation normalized to the maximum value of the entire two dimensional distribution for this EJ-200 tile. 
The profiles are along $2~\text{mm}$ horizontal and vertical bands that either pass through the tile center or are offset by $8~\text{mm}$ from the tile center in x or y.
We observed uniform (within statistical uncertainty) light yield in bands located $8~\text{mm}$ away from the dimple center and $7~\text{mm}$ away from the side edges of the reference tile.

\begin{figure}[ht]
\centering
{
\includegraphics[scale=0.37]{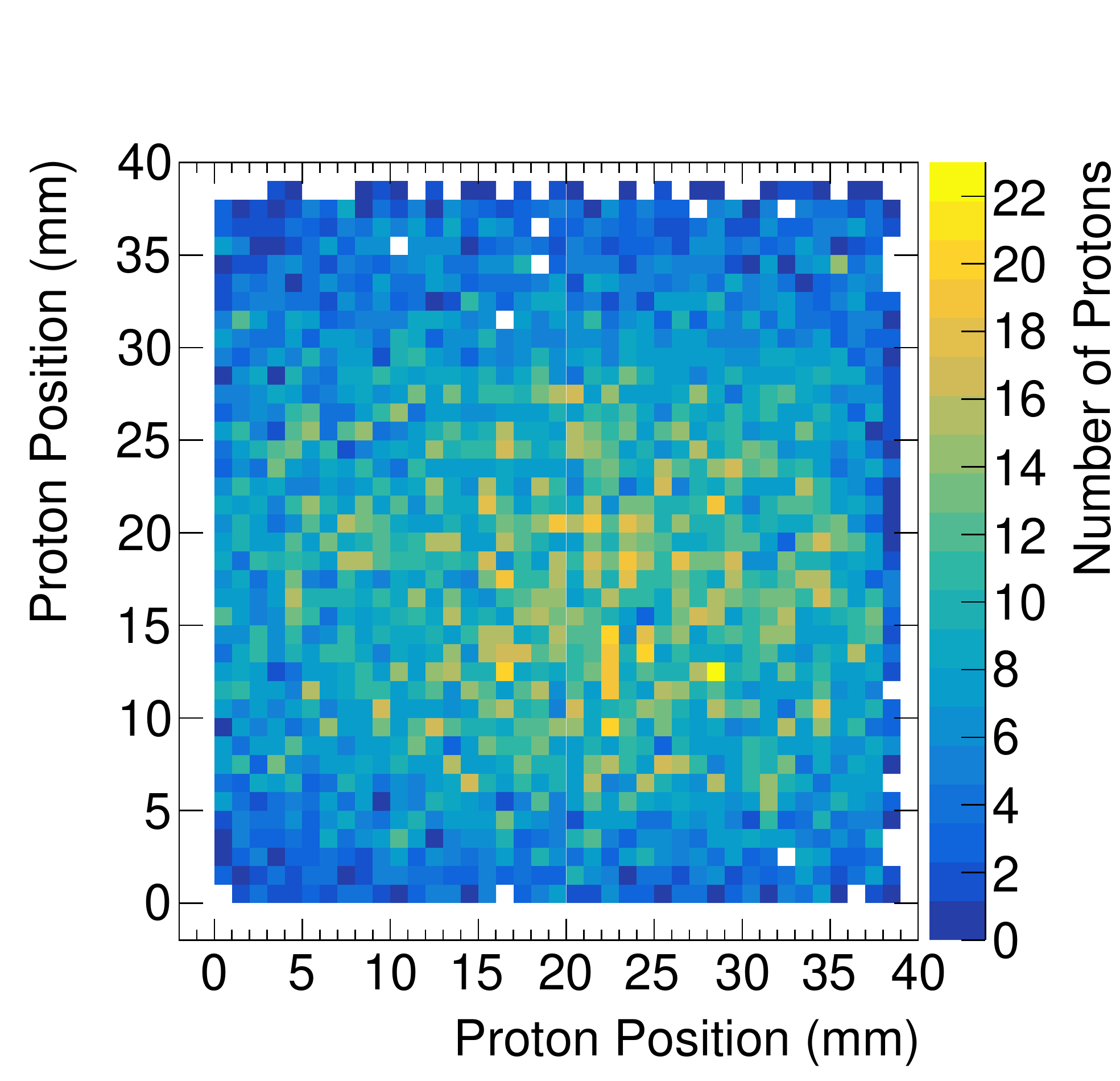}
\includegraphics[scale=0.37]{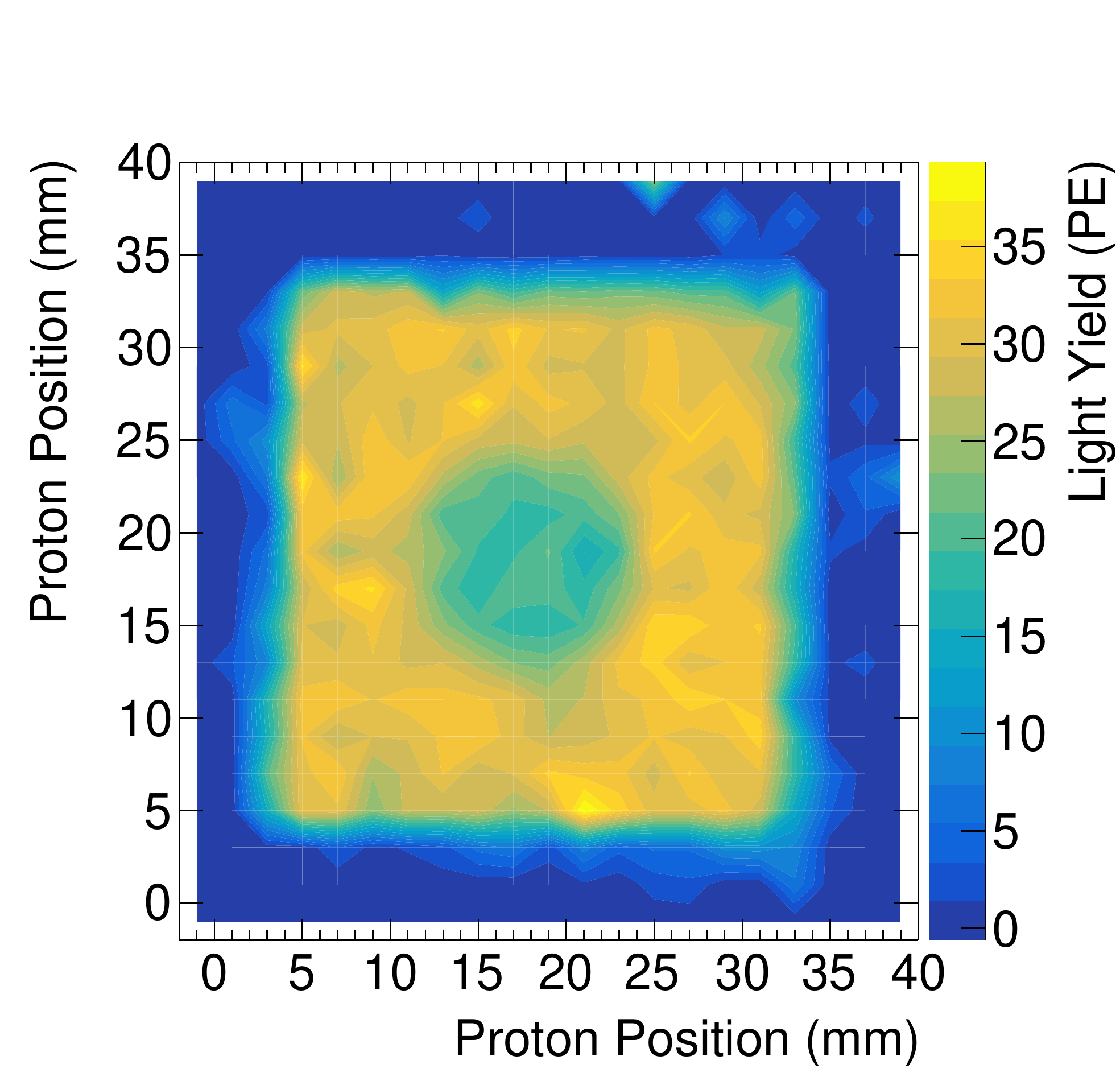}
\leftline{ \hspace{3.7cm}{\bf (a)} \hfill\hspace{3.7cm} {\bf (b)} \hfill}
\caption{\label{fig:uniformity} 
The a) beam intensity  and b)  average light yield  distribution profile 
as a function of  proton position in the upstream tracker. The light yield in b) is that of a SCSN-81 tile with large dimple. 
The reduced light yield at the dimple is clearly seen. }
}
\end{figure}

\begin{figure}[ht]
\centering
{
\includegraphics[scale=0.37]{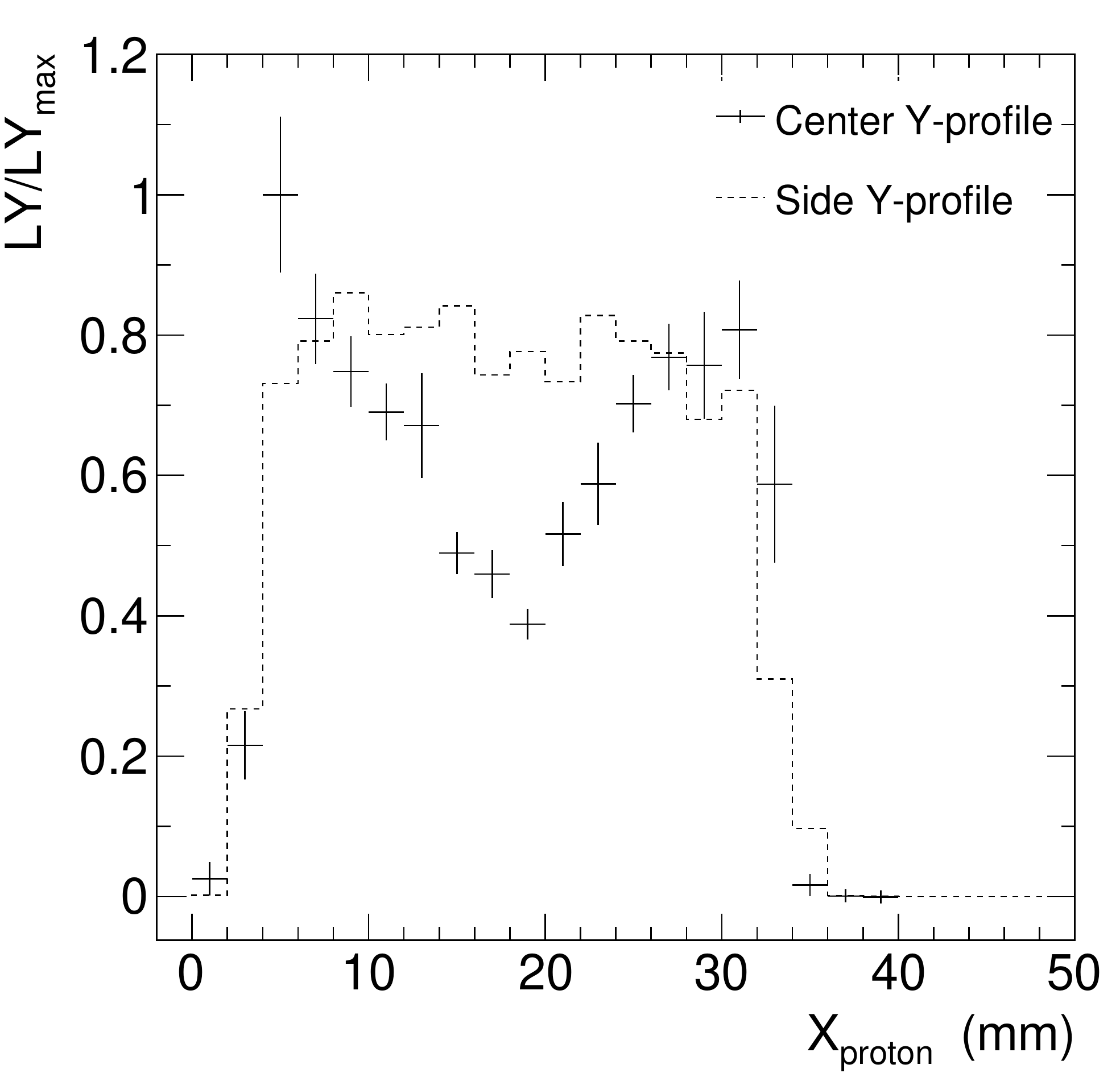}
\includegraphics[scale=0.37]{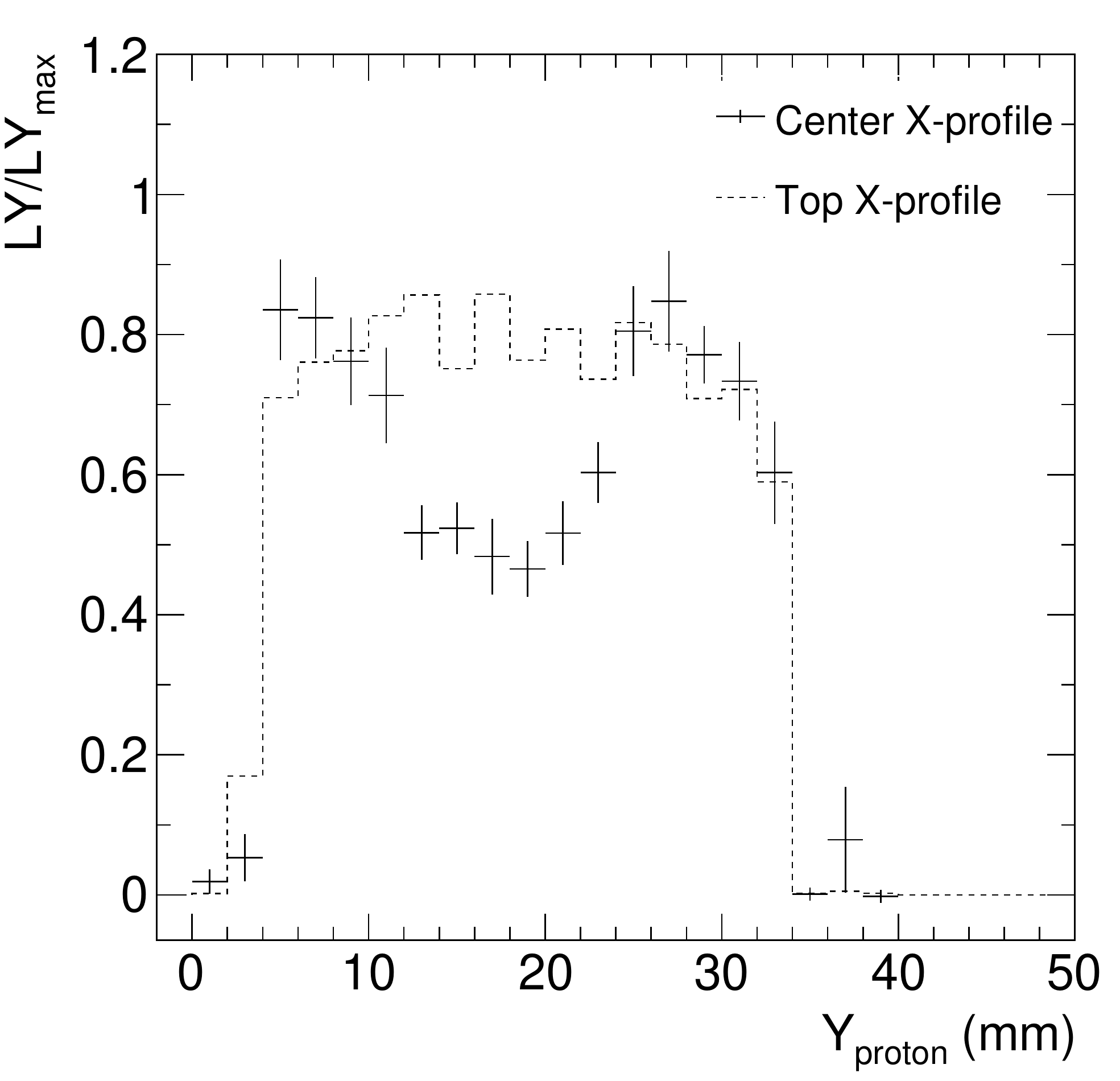}
\leftline{ \hspace{3.7cm}{\bf (a)} \hfill\hspace{3.7cm} {\bf (b)} \hfill}
\caption{\label{fig:uniformity_profiles} 
 Normalized light yield distribution profiles for the tile with an increased dimple size shown in Figure~\ref{fig:uniformity} along 
 a) $2~\text{mm}$ y-bands at the tile center and $8~\text{mm}$ to the right side of the dimple area and b) $2~\text{mm}$ x-bands at the tile center and $8~\text{mm}$ 
 to the top side of the dimple area. Points with error bars correspond to center profiles, while dashed lines show
 the off-center profiles.}
}
\end{figure}

To validate the performance of our experimental equipment, we prepared a SCSN-81 tile with a large dimple machined into it.  This configuration was expected to exaggerate the effect of the dimple on the uniformity of the tile response.
Figure~\ref{fig:uniformity}
shows the beam intensity profile and the average light yield distribution as a function of the proton position in the upstream tracker plane for this tile. 
The depression of the light yield in the tile center is clearly evident and this corresponds to the dimple in the scintillator material. 
Fig.~\ref{fig:uniformity_profiles} shows profiles of light yield distributions, normalized to the maximum value of the intensity profile for this SCSN-81 tile. 
Points with error bars correspond to profiles of SiPM signals
measured in vertical or horizontal bands of $2~\text{mm}$ width, spanning the tile center. Dashed lines show profiles in bands of $2~\text{mm}$ width that are $8~\text{mm}$ offset from the tile center in x or y.

\begin{figure}[ht]
\centering
{
\includegraphics[width=.7\textwidth]{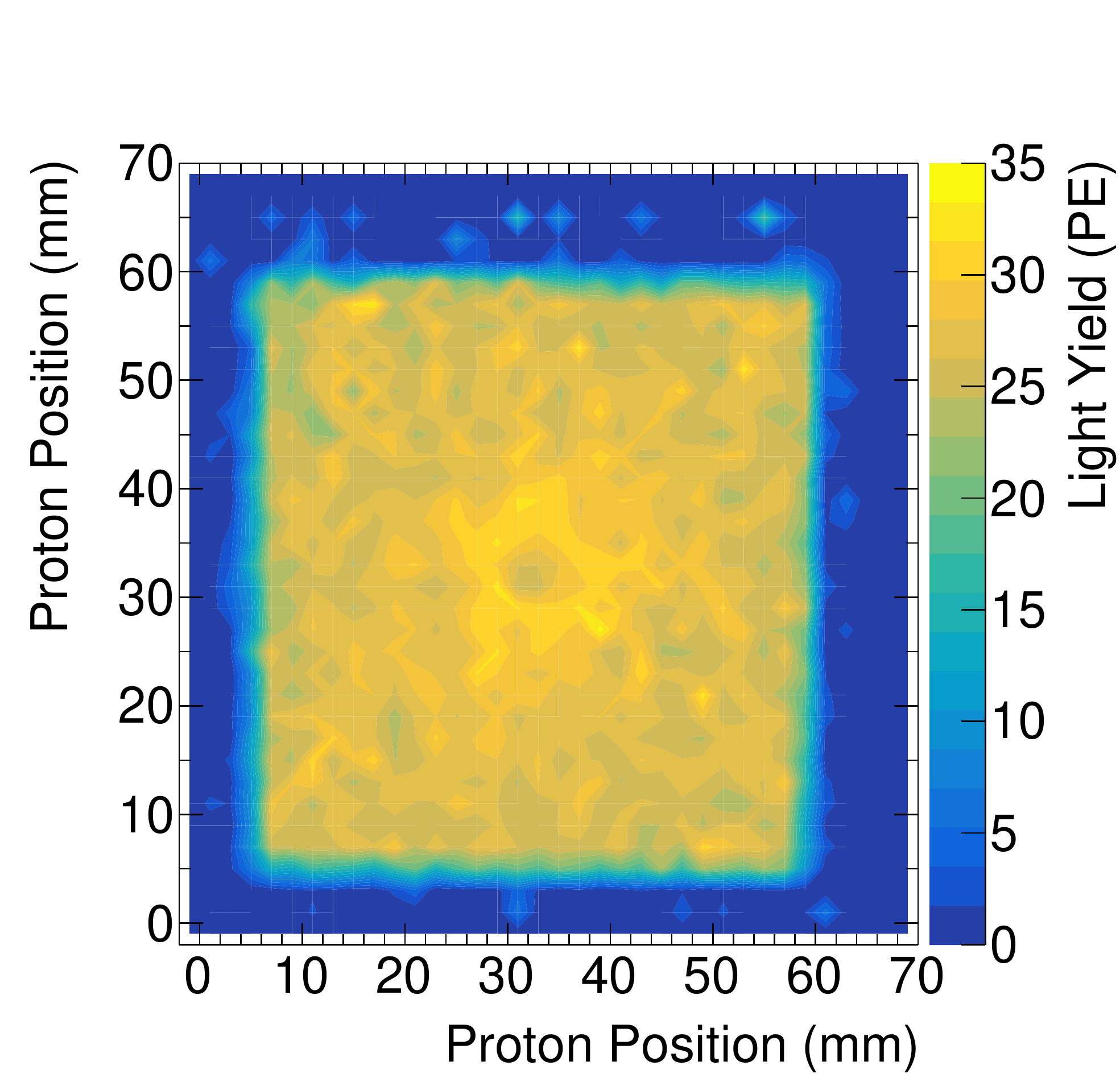}
\caption{\label{fig:5x5EJ200_uniformity} 
The average light yield distribution profile of the $5.5\times5.5~\text{cm}^{2}$ EJ-200 tile as a function of proton position in the upstream tracker.}
}
\end{figure}

\begin{figure}[ht]
\centering
{
\includegraphics[width=.7\textwidth]{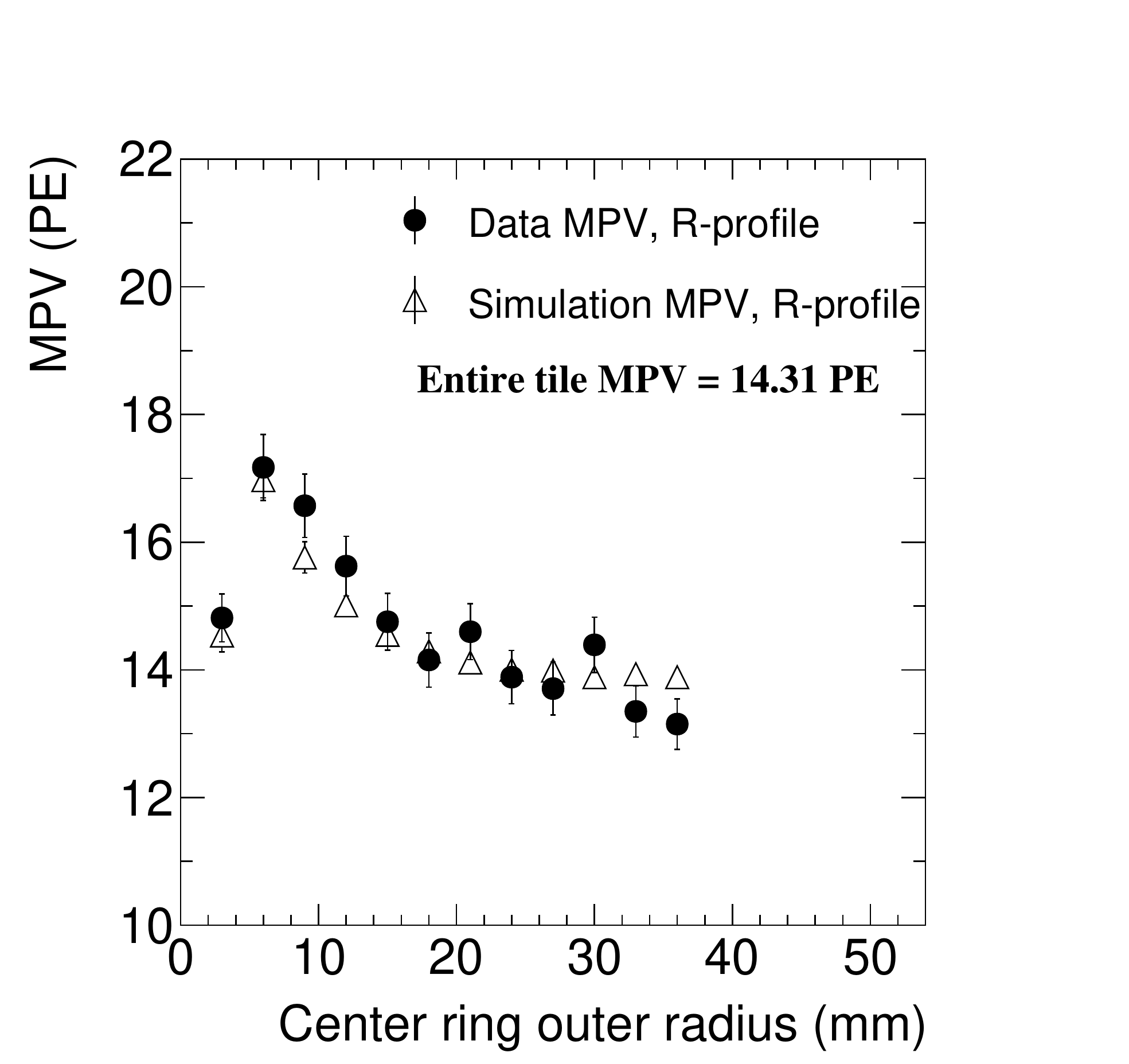}
\caption{\label{fig:5x5EJ200_RadialUniformity} 
The MPV from concentric annular regions centered at the tile's dimple (black points) for the $5.5\times5.5~\text{cm}$ EJ-200 tile. The error bars are $\pm3\%$ systematic, which dominate the statistical errors in the fits. Simulated results are indicated by open triangles.}
}
\end{figure}

We show in Fig.~\ref{fig:5x5EJ200_uniformity} the average light yield distribution as 
a function of proton position for a $5.5\times5.5~\text{cm}^{2}$ EJ-200 tile.
For the same tile, we show the light yield for annular regions centered on the dimple in Fig.~\ref{fig:5x5EJ200_RadialUniformity}.
We observe an increase of the light yield directly in the region outside the dimple.


\begin{figure}[ht]
\centering
{
\includegraphics[scale=0.37]{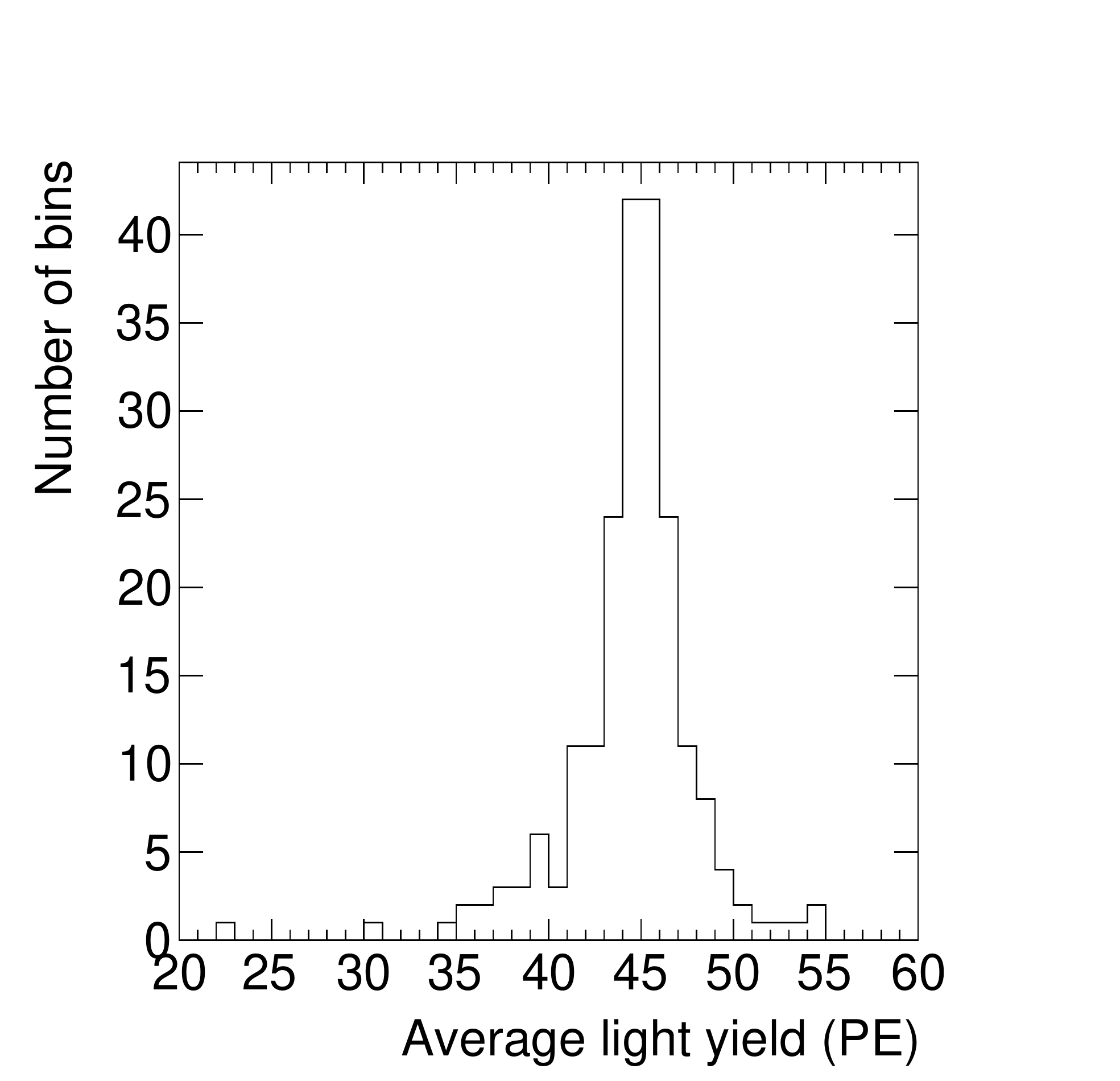}
\includegraphics[scale=0.37]{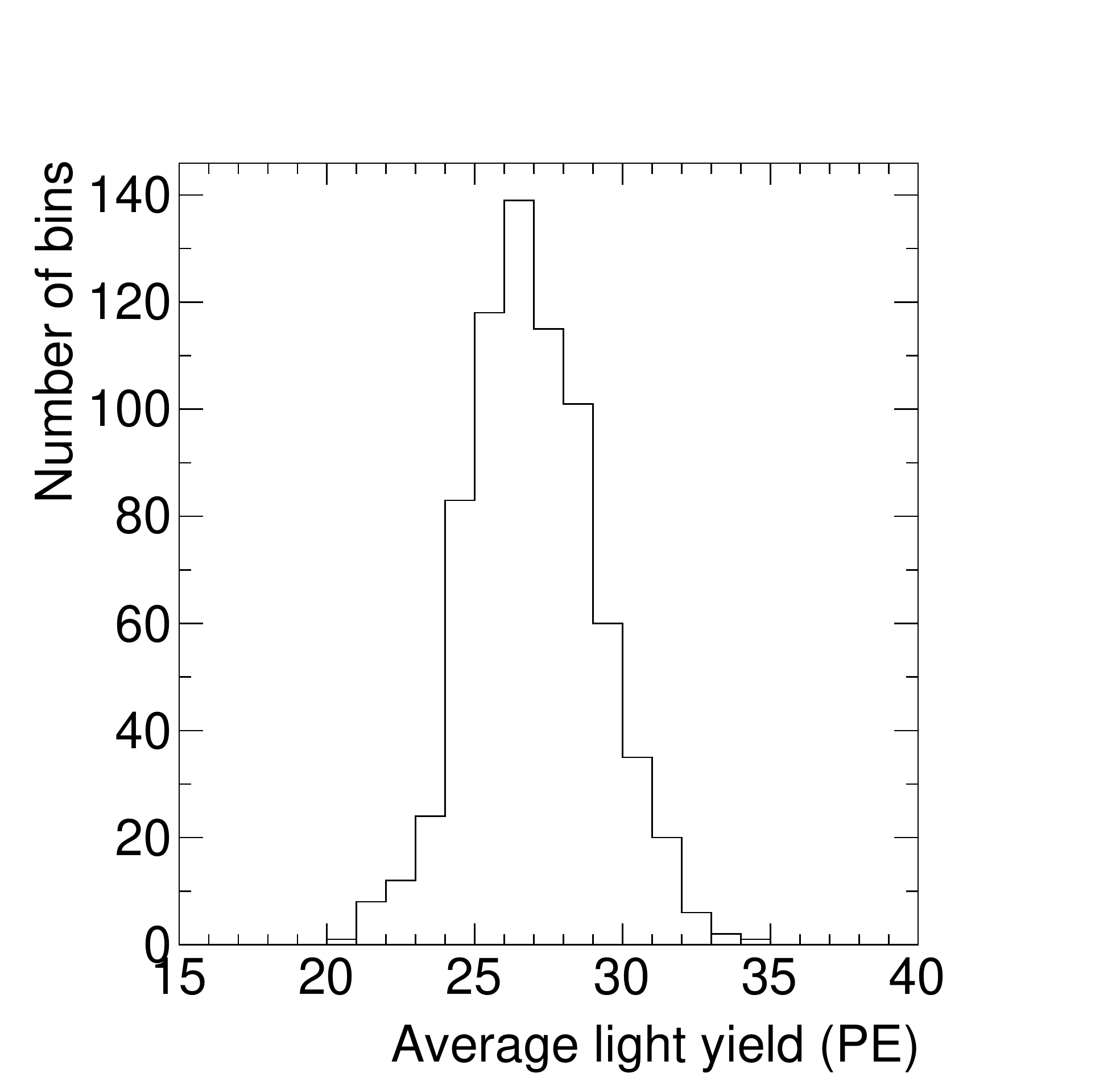}
\leftline{ \hspace{3.7cm}{\bf (a)} \hfill\hspace{3.7cm} {\bf (b)} \hfill}
\includegraphics[scale=0.37]{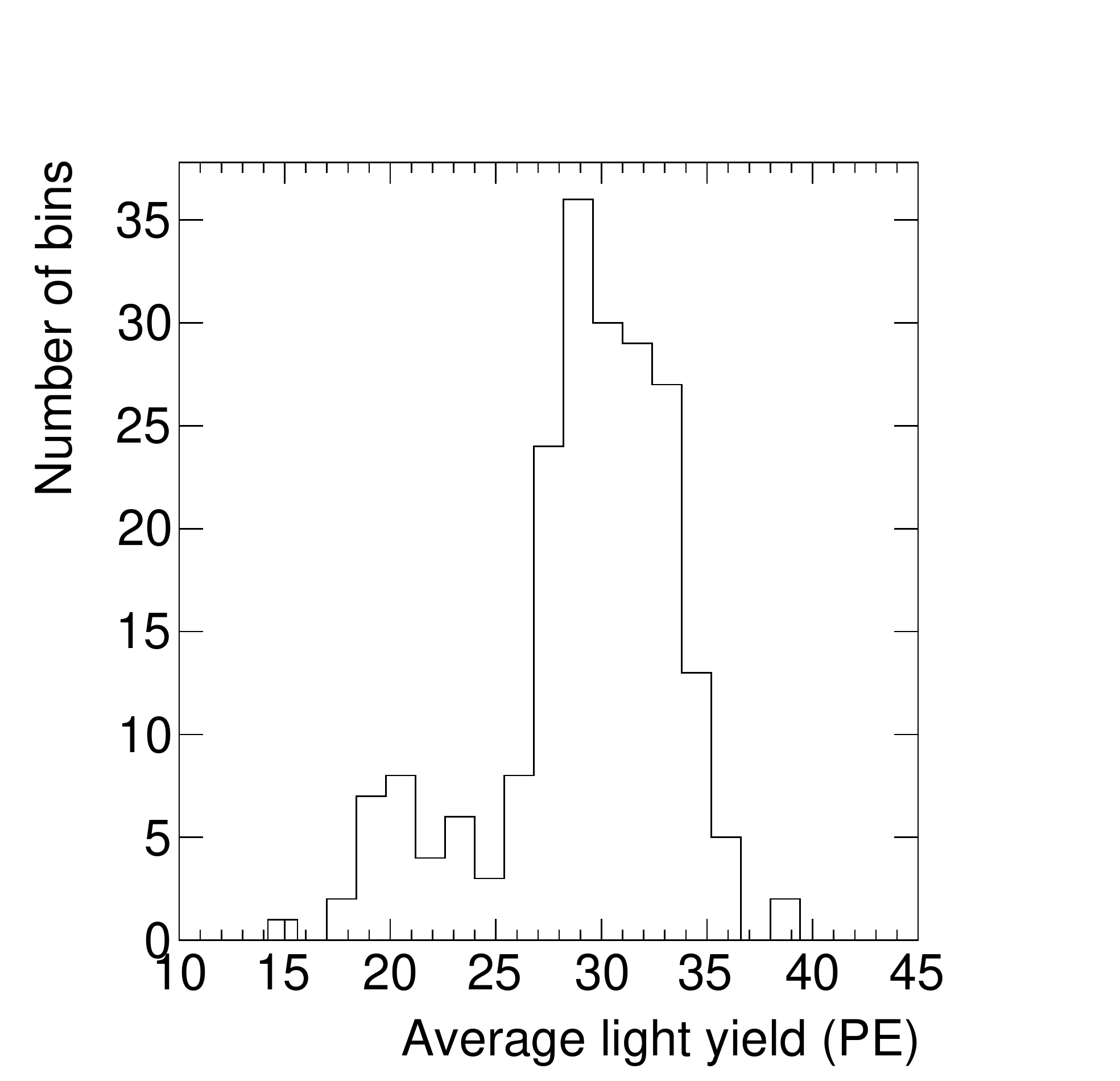}
\leftline{ \hspace{7.1cm}{\bf (c)}}
\caption{\label{fig:AvgLY} 
 The average light yield in $2\times2~\text{mm}$ bins for
 a) the $3\times3~\text{cm}^{2}$ EJ-200 tile, b) the $5.5\times5.5~\text{cm}^{2}$ EJ-200 tile, c) and the SCSN-81 tile with large dimple.}
}
\end{figure}

To quantify a tile's uniformity, we divided a tile into $2\times2~\text{mm}^{2}$ bins, determined the average light yield in each bin, and calculated the $\text{RMS}/\text{mean}$ for all average light yield values in a given tile. The average light yield distributions are shown in Fig.~\ref{fig:AvgLY} for the $3\times3~\text{cm}^{2}$ EJ-200 tile, the $5.5\times5.5~\text{cm}^{2}$ EJ-200 tile, and the SCSN-81 tile with large dimple. 
To establish that the $\text{RMS}/\text{mean}$ distribution provides useful insights on the uniformity of tiles, we performed a simplified simulation of a $3\times3~\text{cm}^{2}$ tile.
The simulated light response across the tile was perfectly uniform, following a Poisson distribution with an average light yield of $30$ PE. 
We simulated 10 events in each $2\times2~\text{mm}^{2}$ bin, the average number of events per bin in data, and compared the result to data by quantifying the non-uniformity as $\frac{\text{RMS}/\text{mean}_{\text{tile}}}{\text{RMS}/\text{mean}_{\text{uniform}}} - 1$.
The non-uniformity of the different sized EJ-200 tiles and the special SCSN-81 tile is listed in Table~\ref{tab:uniformity}.
As expected, the EJ-200 tiles show considerably lower non-uniformity than the special SCSN-81 tile. 

\begin{table}[ht]
  \caption{
    The non-uniformity, $\frac{\text{RMS}/\text{mean}_{\text{tile}}}{\text{RMS}/\text{mean}_{\text{uniform}}} - 1$, of the average light yield for the different sized EJ-200 tiles and the SCSN-81 tile with the large dimple machined into it.
  }
  \centering
  \label{tab:uniformity}
  \begin{tabular}{cc}
Tile &  Non-Uniformity  \\
\hline
 EJ-200 $2.3\times2.3$ & $0.14\pm0.03$ \\
 EJ-200 $3.0\times3.0$ & $0.39\pm0.05$ \\
 EJ-200 $3.4\times3.4$ & $0.61\pm0.05$ \\
 EJ-200 $5.5\times5.5$ & $0.40\pm0.04$  \\
 SCSN-81 $3.0\times3.0$ & $1.58\pm0.11$ \\
\hline
  \end{tabular}
\end{table}

\subsection{Light yield as a function of tile area}
\begin{figure}[ht]
\centering 
\includegraphics[trim={0.1cm 0.1cm 0.1cm 0.1cm}, clip, width=.9\textwidth]{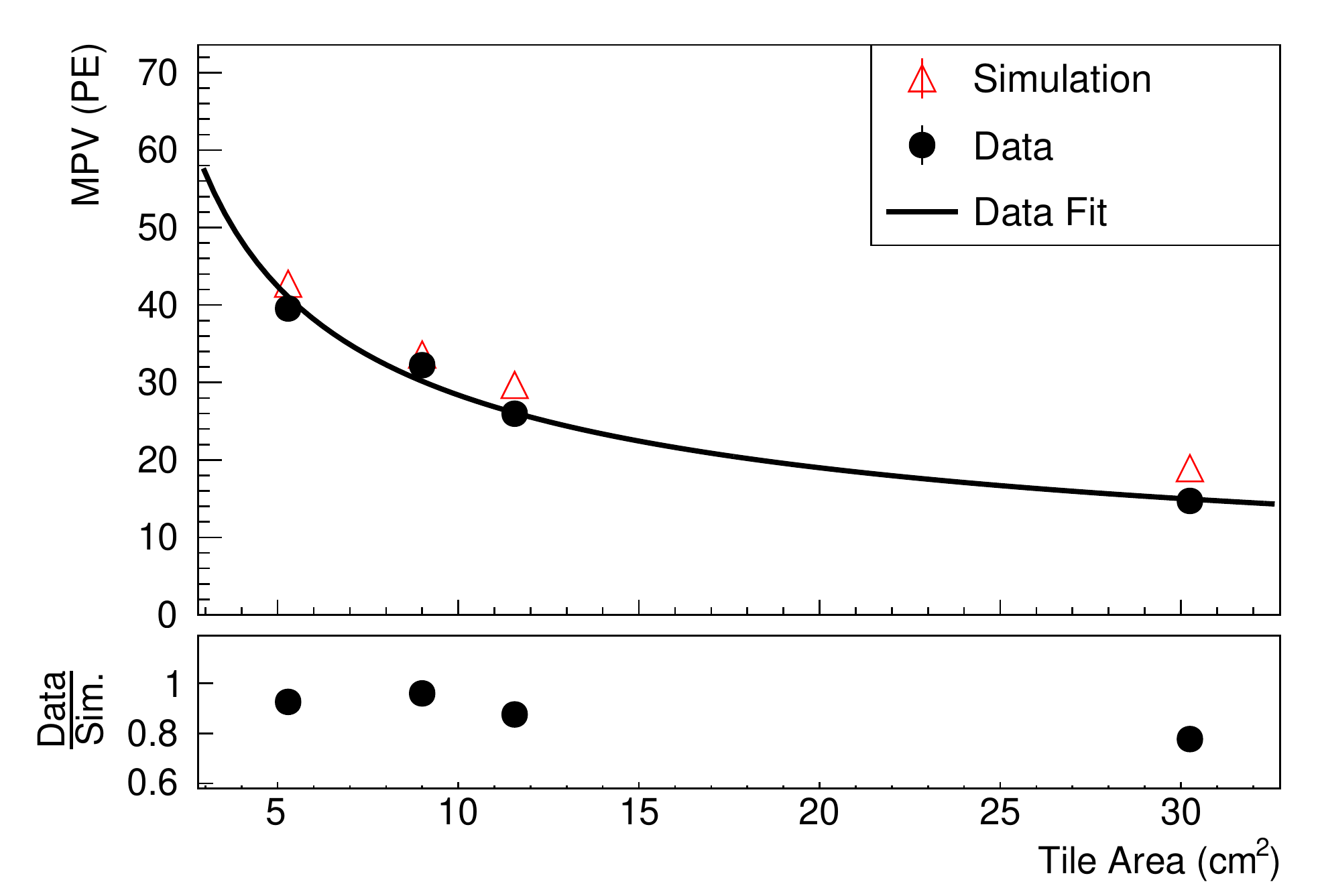}
\caption{\label{fig:MPVvsTileArea} The light yield reported as MPV in data and simulation as a function of tile area for a $3~\text{mm}$ thick, 
wrapped in ESR, EJ-200 scintillator tile. A systematic uncertainty due to reproduciblitiy of optical coupling is estimated to be $3.0\%$, 
and is plotted but smaller than the data points. The black curve is the fit to data. }
\end{figure}

The light yield was measured for different sized square tiles composed of $3~\text{mm}$ thick, EJ-200 scintillator wrapped in ESR. 
Tile sizes ranged from $2.3\times2.3$ -- $5.5\times5.5~\text{cm}^{2}$. The results are compared to simulation and shown in Fig~\ref{fig:MPVvsTileArea}. 
A $\chi^{2}$ fit to data is performed with the function $p_{0}\times(\mathrm{Tile~Area}/9~\text{cm}^2)^{p_{1}}$, where $p_{0}$ and $p_{1}$ are parameters of the fit and fitted values of $p_{0}=30.16\pm0.48$ PE and $p_{1}=-0.58\pm0.02$ are obtained. 
The light yield is roughly inversely proportional to the square root of the tile area.
\subsection{Light yield as a function of tile thickness}
\begin{figure}[ht]
\centering
\includegraphics[width=.47\textwidth]{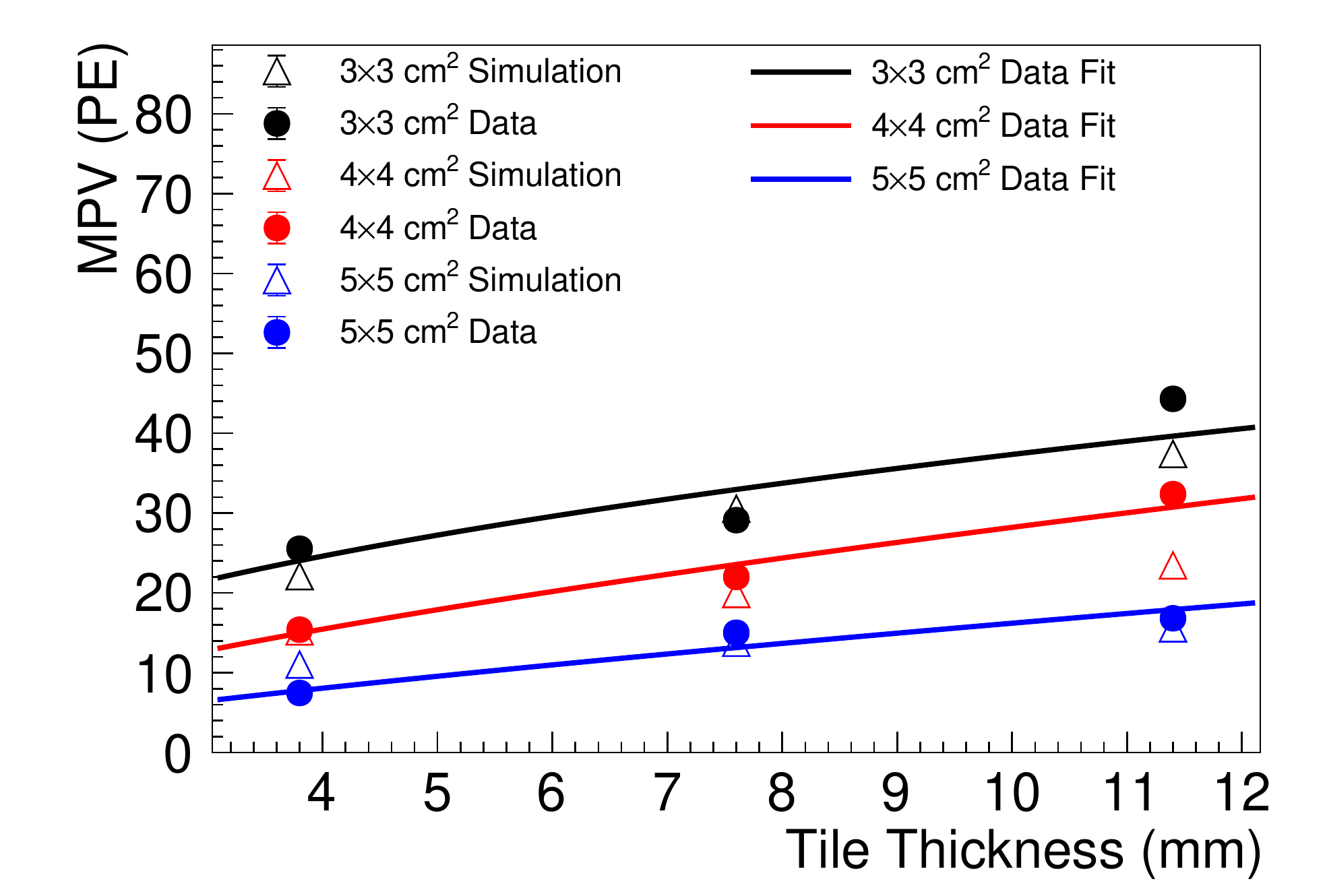} 
\qquad
\includegraphics[width=.47\textwidth]{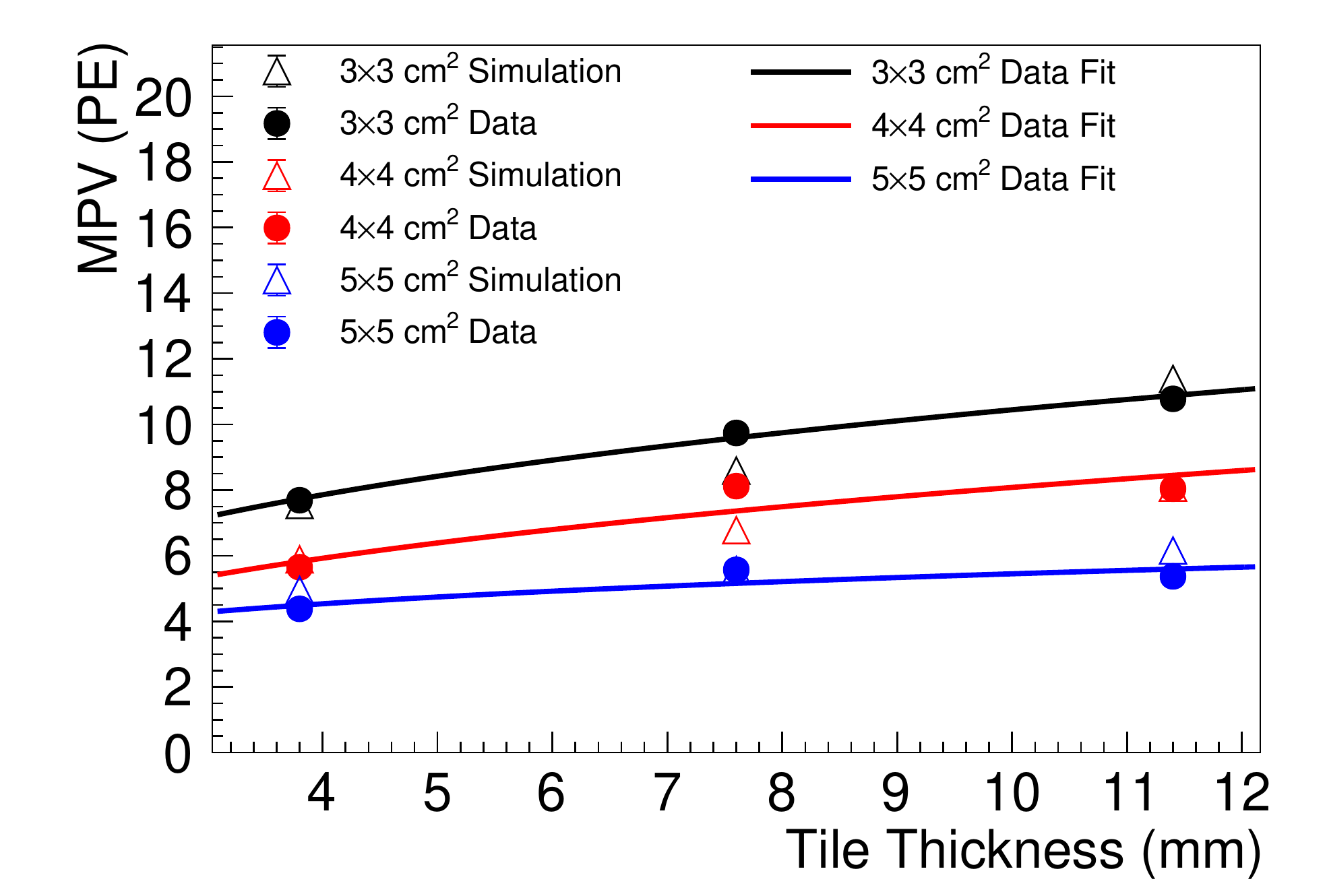} 

\caption{\label{fig:MPVvsTileThickness} The light yield reported as MPV in data and simulation as a function of tile thickness for $3\times3$, $4\times4$, and $5\times5~\text{cm}^{2}$ square SCSN-81 tiles, wrapped in ESR (left) or Tyvek\textsuperscript{\textregistered} (right). The different colored curves are the fits to data.} 
\end{figure}

The light yield was measured for tiles of different thicknesses wrapped in either ESR or Tyvek\textsuperscript{\textregistered}. 
This was done by stacking $3.8~\text{mm}$ thick, SCSN-81 tiles with a thin layer of Bicron BC-630 optical grease in between to simulate a solid tile. 
Care was taken to avoid bubbles in the optical grease coupling. The edges of the tiles were diamond fly-cut to polish them. 
Sets of three square tiles with dimensions $3\times3$, $4\times4$, and $5\times5$ $\text{cm}^{2}$ were prepared. 
From each set, one tile had a dimple machined in it and the other two tiles had unmachined faces. 
The results are compared to simulation and shown in Fig~\ref{fig:MPVvsTileThickness}. 
The discrepancies between data and simulation are caused by imperfections from the optical grease which were not taken into account in the simulation.
A $\chi^{2}$ fit is performed to the light yield
measured in data of each tile thickness with the function $p_{0}\times(\mathrm{Tile~Thickness}/3.8~\text{mm})^{p_{1}}$, where $p_{0}$ and $p_{1}$ are parameters of the fit and the fitted values are shown in Table~\ref{tab:fitValues}. In all cases the increase in light yield was less than linear in tile thickness.

The light yield difference between tiles wrapped with ESR vs. Tyvek\textsuperscript{\textregistered} is shown in Table~\ref{tab:TyvekESR}, where the ratio of MPV from ESR wrapped tiles to Tyvek\textsuperscript{\textregistered} wrapped tiles for each sample size is listed. In all cases, ESR wrapped tiles demonstrate much higher light yields than tiles wrapped in Tyvek\textsuperscript{\textregistered}.

\begin{table}[ht]
  \caption{
    The fit values of $p_{0}$ and $p_{1}$ in the $p_{0}\times(\mathrm{Tile~Thickness}/3.8~\text{mm})^{p_{1}}$ fit to the different sized and wrapped tiles.
  }
  \centering
  \label{tab:fitValues}
  \begin{tabular}{cccc}
SCSN-81 tile area, mm &  Wrapping & $p_{0}$ fit values & $p_{1}$ fit values \\
\hline
 $30.0\times30.0$ & ESR & $24.03\pm0.74$ & $0.45\pm0.04$ \\
 $30.0\times30.0$ & Tyvek\textsuperscript{\textregistered} & $7.73\pm0.22$ & $0.31\pm0.04$ \\
 $40.0\times40.0$ & ESR & $14.95\pm0.44$ & $0.66\pm0.04$ \\
 $40.0\times40.0$ & Tyvek\textsuperscript{\textregistered} & $5.82\pm0.16$ & $0.34\pm0.04$ \\
 $50.0\times50.0$ & ESR & $7.76\pm0.21$ & $0.76\pm0.04$ \\
 $50.0\times50.0$ & Tyvek\textsuperscript{\textregistered} & $4.49\pm0.13$ & $0.20\pm0.04$ \\
\hline
  \end{tabular}
\end{table}

\begin{table}[ht]
  \caption{
     The ratio of MPV for ESR divided by  Tyvek\textsuperscript{\textregistered} wrapped tiles.
  }
  \centering
  \label{tab:TyvekESR}
  \begin{tabular}{cc}
SCSN-81 tile dimensions, mm & ESR/Tyvek\textsuperscript{\textregistered} light yield MPV ratio \\
\hline
 $30.0\times30.0~\times3.8$  & $3.32 \pm 0.30$ \\
 $30.0\times30.0~\times7.6$  & $2.99 \pm 0.27$ \\ 
 $30.0\times30.0~\times11.4$ & $4.11 \pm 0.37$ \\
 $40.0\times40.0~\times3.8$  & $2.72 \pm 0.25$ \\
 $40.0\times40.0~\times7.6$  & $2.63 \pm 0.24$ \\
 $40.0\times40.0~\times11.4$ & $4.03 \pm 0.37$ \\
 $50.0\times50.0~\times3.8$  & $1.71 \pm 0.16$ \\
 $50.0\times50.0~\times7.6$  & $2.65 \pm 0.24$ \\
 $50.0\times50.0~\times11.4$ & $3.15 \pm 0.29$ \\ 
\hline
  \end{tabular}
\end{table}

\subsection{Light yield as a function of hole size in ESR reflective wrapper}

\begin{figure}[ht]
\centering 
\includegraphics[trim={0.1cm 0.1cm 0.1cm 0.1cm}, clip, width=.8\textwidth]{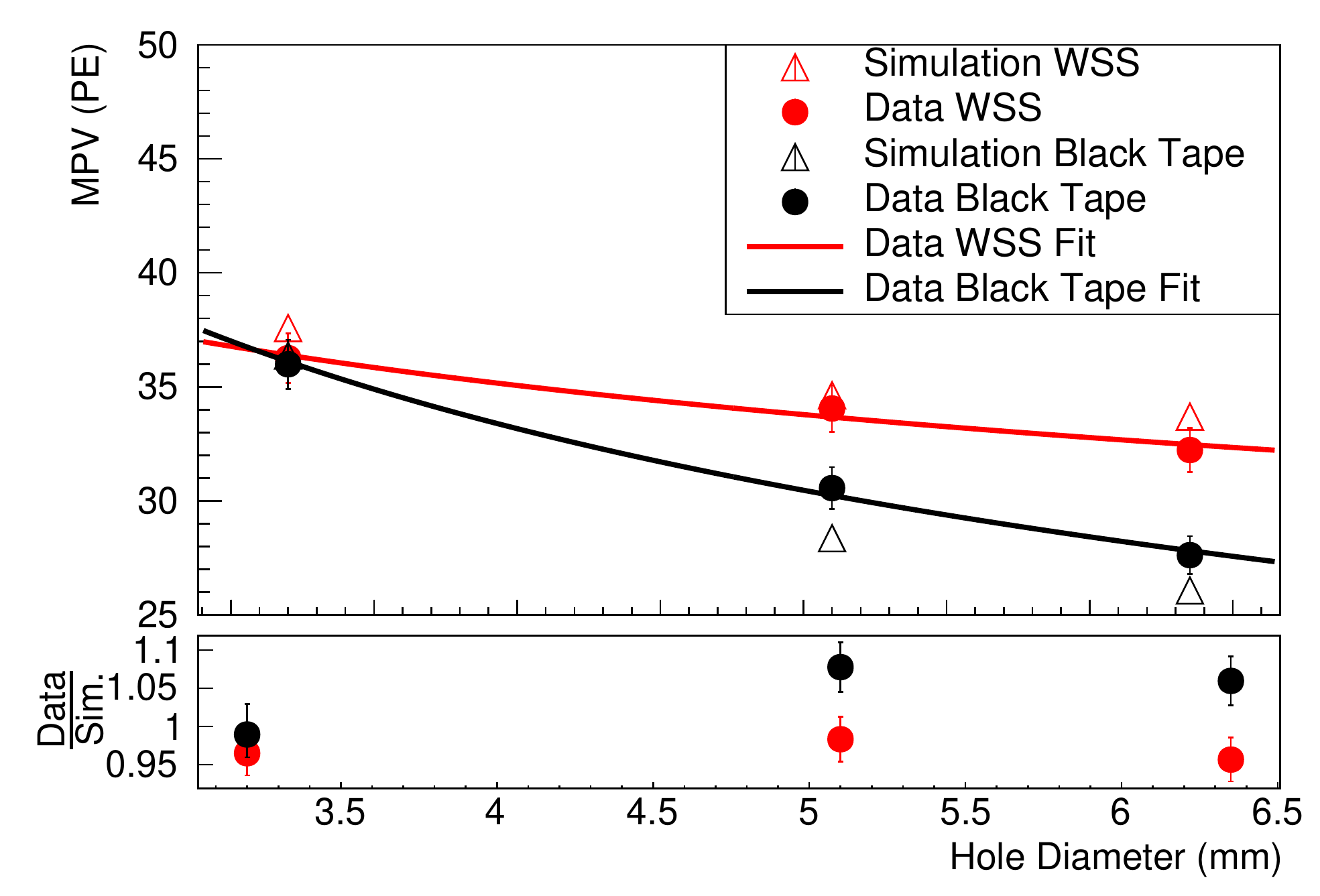}

\caption{\label{fig:MPVvsHoleSize} The MPV of the light yields in data and simulation as a function of hole size are shown 
for SiPMs sitting on top of a white silkscreened backplate and a black tape backplate. 
 (S13360 SiPM, EJ-200 tile.)}
\end{figure}

The light yield as a function of hole size in the ESR reflector was measured for a $3\times3~\text{cm}^{2}$ EJ200 tile wrapped in ESR. 
One expects that as the hole diameter in the reflective foil decreases, the light yield should increase, since fewer photons escape through the gap between SiPM and wrapper. 
Likewise, it is expected that if the white silkscreen is changed to a non-reflective surface, the light yield should decrease, 
since fewer photons reflect back into the tile through the gap. The light yield was studied for cases of a white 
silk screened backplate, and holes in the reflector of $3.2~\text{mm}$, $5.1~\text{mm}$, and $6.35~\text{mm}$ diameter. For these hole sizes, 
the white backplate was compared to data taken with the white backplate covered with black tape, 
which is a good approximation to a nonreflective surface. 
The results are compared to simulation and shown in Fig.~\ref{fig:MPVvsHoleSize}.
A $\chi^{2}$ fit to data is performed with the function $p_{0}\times(\mathrm{Hole~Diameter}/3.2~\text{mm})^{p_{1}}$ where $p_{0}$ and $p_{1}$ are parameters of the fit and fitted values of $p_{0}=36.53\pm1.56$ PE and $p_{1}=-0.17\pm0.06$ for WSS and $p_{0}=16.44\pm0.70$ PE and $p_{1}=-0.38\pm0.06$ for black tape are extracted.
In accordance with our expectations, larger holes had lower light yields, and the effect is stronger for the black backplate.

\section{Systematics}
\subsection{Reproducibility of optical coupling}
\label{sec:OpticalCoupling}
A systematic uncertainty was calculated for the reproducibility 
of the optical coupling between tile and SiPM by comparing the MPV across 
nominally identical measurements of the same tile with the same wrapping.
A typical reproducibility of $3\times3~\text{cm}^{2}$ tiles was found to be $\pm3\%$, 
derived from the root mean square deviation from the mean of six measurements of the same tile. 
\subsection{Misalignment of SiPM relative to the tile dimple}
We studied intentional misalignment of the SiPM relative to its nominal location in the x-y plane, 
and also in its depth into dimple (z direction). 
For a SiPM misaligned by $1.7~\text{mm}$, about four times larger than our alignment uncertainty, in the x-y~plane, the difference in the MPV response was found at the level of $2.7\%$, an effect that was small compared to our overall reproducibility uncertainty.
To study the misalignment in the z-direction, we moved the SiPM $1~\text{mm}$ away from the 
tile and found the MPV decreased by $36.6\%$. Although this is a significant effect, we believe it is not important for our studies due to the design of the mechanics of our test stand that naturally provided consistent z placement. Additionally, the optical coupling reproducibility measurement, Sec~\ref{sec:OpticalCoupling}, includes any residual effect of z misalignment.
Our tile-to-SiPM mounting was very reproducible for most tiles. 
\subsection{SiPM temperature dependence}
The temperature of the SiPM was measured using a PT10000 temperature sensor placed on the PCB close to the SiPM. 
Using this sensor, we observed a temperature variation between 26 and 28$^{\circ}$C for all runs.  
For our SiPMs and our choice of over-voltage, the expected change in the number of detected photons due to temperature variation is less than 1\%. 
The systematic uncertainty in optical coupling covers this uncertainty. 
\section{Summary}
A setup to study the responses of different scintillator tile geometries and materials has been installed at the Fermilab Test Beam Facility.
This was used to collect and analyze data in a January-February 2020 test beam run.   
The light yield and uniformity was measured for various values of the tile size,  tile thickness, type of the reflective wrapper, and the hole diameter in the wrapper. 
ESR was found to be substantially more reflective than Tyvek\textsuperscript{\textregistered}, 
with a ratio of light yield of roughly $2$ to $4$ times, depending on other tile parameters. 
The light yield of a tile was measured as approximately inversely proportional to the square root of the tile area. 
The light yield was observed to increase much less than linearly with thickness, with the exact amount depending on tile size and wrapping material.  
Simulation was developed that agrees well with the light yield and uniformity measured in test beam data across the different tiles. 

\acknowledgments
Work supported by the Fermi National Accelerator Laboratory, managed and operated by Fermi Research Alliance, LLC under Contract No. DE-AC02-07CH11359 with the U.S. Department of Energy. The U.S. Government retains and the publisher, by accepting the article for publication, acknowledges that the U.S. Government retains a non-exclusive, paid-up, irrevocable, world-wide license to publish or reproduce the published form of this manuscript, or allow others to do so, for U.S. Government purposes. Work also supported by the US-DOE Office of Science (High Energy Physics) under Award Number  DE-SC0011845 and DE‐SC0010072. Additional support provided by the University of Maryland Physics Department.

\end{document}